\documentclass[12pt,preprint]{aastex} 
\usepackage{graphics,graphicx} 
\usepackage{natbib}

\newcommand{\skipthis}[1]{}

\def\nh3{$\rm{NH_3}$}
\def\NH3{$\rm{NH_3}$} 
 
\def\msun{M$_\odot$}
 
\def\lsun{L$_\odot$} 
\def\kms-1{km~s$^{-1}$}
\def\h2o{$\rm{H_2O}$} 
\def\h2{$\rm{H_2}$} 
\def\CM2{$\rm{cm^{-2}}$}
\def\cm3{$\rm{cm^{-3}}$}

\def\n2h+{N$_2$H$^+$}
\def\ch3oh{CH$_3$OH}

\newcommand{\lsim}{${\raisebox{-.9ex}{$\stackrel{\textstyle<}{\sim}$}}$ }

\begin{document}


\title{Magnetic Fields and Massive Star Formation }

\author{Qizhou Zhang\altaffilmark{1}, Keping Qiu\altaffilmark{2}, Josep M. Girart\altaffilmark{3}, Hauyu (Baobab) Liu\altaffilmark{4},  Ya-Wen Tang\altaffilmark{4}, Patrick M. Koch\altaffilmark{4}, Zhi-Yun Li\altaffilmark{5}, Eric Keto\altaffilmark{1}, Paul T. P. Ho\altaffilmark{1,4}, Ramprasad Rao\altaffilmark{4}, Shih-Ping Lai\altaffilmark{6,4}, Tao-Chung Ching\altaffilmark{1,6}, Pau Frau\altaffilmark{7}, How-Huan Chen\altaffilmark{1}, Hua-Bai Li\altaffilmark{8}, Marco Padovani\altaffilmark{9}, Sylvain Bontemps\altaffilmark{10}, Timea Csengeri\altaffilmark{11}, Carmen Ju\'arez\altaffilmark{3}}

\altaffiltext{1}{Harvard-Smithsonian Center for Astrophysics, 60 Garden Street,
Cambridge MA 02138, USA. E-mail: qzhang@cfa.harvard.edu}

\altaffiltext{2}{School of Astronomy and Space Science, Nanjing University, 22 Hankou Road, Nanjing 210093, China}

\altaffiltext{3}{Institut de Ci{\`e}ncies de l'Espai, (CSIC-IEEC), Campus UAB, Facultat de Ci{\`e}ncies, C5p
2, 08193 Bellaterra, Catalonia }

\altaffiltext{4}{Academia Sinica Institute of Astronomy and Astrophysics, P. O. Box 23-141, Taipei, 106 Taiwan }

\altaffiltext{5}{Department of Astronomy, University of Virginia, P. O. Box
400325, Charlottesville, VA 22904, USA }

\altaffiltext{6}{Institute of Astronomy and Department of Physics, National Tsing Hua University, 101 Section 2 Kuang Fu Road, Hsinchu 30013, Taiwan}

\altaffiltext{7}{Observatorio Astron\'{o}mico Nacional, Alfonso XII, 3 E-28014 Madrid, Spain }

\altaffiltext{8}{Department of Physics, The Chinese University of Hong Kong }

\altaffiltext{9}{Laboratoire de Radioastronomie Millim{\'e}trique, UMR 8112 du CNRS, {\'E}cole Normale Sup{\'e}rieure et Observatoire de Paris,
24 rue Lhomond, 75231 Paris Cedex 05, France }

\altaffiltext{10}{OASU/LAB-UMR5804, CNRS, Universit\'{e} Bordeaux 1, 33270 Floirac, France }

\altaffiltext{11}{Max Planck Institute for Radioastronomy, Auf dem H\"{u}gel 69, 53121 Bonn, Germany }

\email{qzhang@cfa.harvard.edu}


\begin{abstract}

Massive stars ($M > 8$ \msun) typically form in parsec-scale molecular clumps that collapse and fragment, leading to the birth of a cluster of stellar objects. We investigate the role of magnetic fields in this process through dust polarization at 870 $\mu$m obtained with the Submillimeter Array (SMA). The SMA observations reveal polarization at scales of $\lsim$ 0.1 pc. The polarization pattern in these objects ranges from ordered hour-glass configurations to more chaotic distributions. By comparing the SMA data with the single dish data at parsec scales, we found that magnetic fields at dense core scales are either aligned within $40^\circ$ of or perpendicular to the parsec-scale magnetic fields. This finding indicates that magnetic fields play an important role during the collapse and fragmentation of massive molecular clumps and the formation of dense cores. We further compare magnetic fields in dense cores with the major axis of molecular outflows. Despite a limited number of outflows, we found that the outflow axis appears to be randomly oriented with respect to the magnetic field in the core. This result suggests that at the scale of accretion disks ($\lsim 10^3$ AU), angular momentum and dynamic interactions possibly due to close binary or multiple systems dominate over magnetic fields. With this unprecedentedly large sample massive clumps, we argue on a statistical basis that magnetic fields play an important role during the formation of dense cores at spatial scale of 0.01 - 0.1 pc in the context of massive star and cluster star formation.

\end{abstract}

\section{Introduction}

Stars are assembled in molecular clouds where over density regions collapse when gravity overcomes the internal pressure.
Magnetic fields and turbulence are the two main forces that counteract gravity and hinder star formation.
For dark globules that normally form low- or intermediate-mass stars, turbulence appears to be sub or 
trans Alfvenic \citep{heyer2008, franco2010}. 
Isolated cores\footnote{This paper follows the nomenclature used in \citet{zhang2009}, and refers 
a {\it cloud} as an entity of molecular gas of $10 - 100$ pc, a molecular {\it clump} as an entity of 1 pc 
that forms massive stars with a population of lower mass stars, and {\it dense cores} as an entity of 0.01 to 0.1 pc
that forms one or a group of stars.}
embedded in these dark clouds are found to contain subsonic turbulence \citep{myers1983,goodman1998}, 
whose role in preventing gravitational collapse may be limited.
Therefore, magnetic fields are proposed to play a dominant role in controlling isolated low-mass star formation \citep{mouschovias1976,shu1977,shu1987}, although this picture has been challenged recently \citep[see][for a review]{crutcher2012}.

Massive stars are found predominantly in regions clustered with lower mass stellar objects \citep{lada2003}. Massive star formation takes place in parsec-scale molecular clumps embedded in giant molecular clouds that collapse and fragment into $\sim$ 0.1 pc dense cores. It is well documented observationally that cluster forming molecular clumps are turbulent with typical FWHM linewidths of $2 - 5$ \kms-1 \citep{jijina1999,pillai2006b,csengeri2011,sanchez2013a,lu2014}, and are massive with typical masses of $10^3$ \msun \citep{beuther2002a,rathborne2006}. Direct measurements of magnetic fields through Zeeman splitting reveal the line-of-sight component in several massive star forming clumps, including DR 21(OH) (0.36 mG) \citep{falgarone2008}, implying an important role of magnetic fields in the dynamical evolution of the clump. Numerical simulations of magnetized, turbulent clouds
have found that filamentary structures form as a result of turbulent driving \citep{ostriker2001,nakamura2008,wang2010,vanloo2014}. In the presence of strong magnetic fields, the filaments appear to be organized with the long axis either perpendicular or parallel to the magnetic field lines \citep{li2013}. In contrast, filaments do not correlate with the direction of magnetic field lines if the field is weak \citep[see][for a recent review]{li2014}.

This paper investigates the role of magnetic fields in massive star formation. We focus our studies to the length scale of $\lsim$ 1 pc, the size of massive molecular clumps. With the spatially resolved images of molecular clumps in dust polarization obtained from the Submillimeter Array, we investigate the role of magnetic fields from 1 pc scale to the scale of dense cores of 0.01 - 0.1 pc, and indirectly further to $\sim$ 1000 AU scale of envelopes and accretion disks of massive protostars. 

We use the polarized emission from dust grains to probe the magnetic field configuration in molecular clouds. In the presence of magnetic fields, elongated grains with paramagnetic properties tend to spin with their angular momentum and grain minor axis aligned with the field direction \citep[e.g.][]{lazarian2007}. Thus, the E vectors of the polarized emission in the millimeter and sub-millimeter wavelengths are perpendicular to the magnetic fields. This technique probes the magnetic field component in the plane of the sky.  Other techniques of probing magnetic fields are not discussed here since they are out of the scope of this paper, but can be found in the recent review article by \citet{crutcher2012}.

One of the best examples supporting the magnetically dominated star formation scenario comes from dust polarization observations of the low-mass dense core surrounding the protostar NGC1333 IRAS4A. These observations reveal a well organized magnetic field threading a flattened dense core \citep{girart2006}. The magnetic field lines are pinched in an hour-glass morphology along the major axis of the dense core, as expected in a collapsing magnetized core. Indeed, a detailed comparison shows that the model of collapsing magnetized cores \citep[e.g.,][]{galli1993, allen2003a} is well matched with observations \citep{goncalves2008,frau2011}. Studies of similar star forming regions have also revealed a pinched magnetic field configuration \citep[e.g.,][]{rao2009,stephens2013,hull2013}. Quantitative analysis from the dispersion of polarization finds the strength of the plane-of-sky component of the magnetic field of several mG, consistent with the picture that magnetic fields are important and perhaps even dominant as compared to turbulence in dense cores of low-mass star formation \citep{girart2006,houde2004}.
 
Studies of dust polarization in massive star forming regions have been carried out with
both single dish telescopes and radio interferometers. Observations with single dish telescopes
such as JCMT or CSO have angular resolutions of 14$''$ to 20$''$ at 850 $\mu$m, or linear scales of 
$0.35 - 0.5$ pc for sources at a distance of 5 kpc, and often reveal ordered dust polarization \citep{matthews2009, dotson2010}, which
implies organized magnetic fields. While the implication of organized magnetic fields is still debated,
the most plausible interpretation is that the fields are dynamically important, thus they are not severely distorted by other dynamic processes in the cloud \citep[e.g.][]{ostriker2001}. Within the 1 pc scale at which massive stars and cluster formation
take place, interferometric observations are required to spatially resolve the morphology of dust
polarization. Limited observations \citep{rao1998, lai2001, cortes2008,girart2009,tang2009a,tang2009b,tang2010,tang2013,liu2013,qiu2013,girart2013,frau2014}
reveal a range for magnetic field morphologies from the well organized hour-glass configuration in G31.41 and W51e2 to
the more chaotic distribution in G5.89. In order to study the role of magnetic fields in massive star formation, we undertook a large observing program using the SMA to map dust polarization in the 345 GHz band in a large sample of massive molecular clumps.
In this first summary paper, we present the survey and a statistical analysis of the dust polarization data. The main questions to be addressed in this paper are: (1) How does the magnetic field configuration vary with spatial scales?
And (2) What is the role of magnetic fields during the fragmentation of molecular clumps and formation of dense cores and massive stars? Detailed studies of individual objects and more quantitative analysis of the sample will
be presented in separate publications \citep[e.g.,][]{liu2013,qiu2013,girart2013}. This paper is organized as follows: Section 2 describes the observations.
Section 3 presents maps of dust polarization and molecular outflows, and comparison of polarization with gas/dust morphology at different spatial scales. In Section 4, we discuss the implications of the results. A summary of the key results is given in Section 5.

\section{Observations}

Polarimetric observations with the SMA\footnote{The Submillimeter
Array is a joint project between the Smithsonian
Astrophysical Observatory and the Academia Sinica Institute of Astronomy and
Astrophysics, and is funded by the Smithsonian Institution and the Academia Sinica.} \citep{ho2004}
of a sample of 21 massive star forming regions were carried out from 2008 through 2012
in the 345 GHz band. The polarimeter on the SMA  uses quarter waveplates that convert the linear polarization
signals to circular ones. A detailed discussion of the system design and performance is provided in \citet{marrone2006, marrone2008}.  Data obtained in 2008 and 2009 have a bandwidth of 2 GHz in each of the lower and upper sideband, separated by an IF frequency of 4 to 6 GHz.
The majority of the observations were carried out during 2011 through 2012 with a 4 GHz bandwidth in each sideband and IF frequencies of 4 to 8 GHz.
The digital correlator was configured to a uniform resolution of 128 channels per 104 MHz chunk, providing a channel width of 0.8 MHz or 0.65 \kms-1 at the observing frequencies. The 345 GHz receivers were tuned to the
CO 3-2 line in the USB. Most of the tracks were observed with 7 or 8 antennas with a zenith opacity at 225 GHz of 0.1 or better.
The majority of the targets were first observed in the compact or subcompact configurations. Sources with strong polarization fluxes  ($> 5\sigma$ levels) were followed up in the extended configuration. The combined visibility data provides a wide range of spatial frequencies from 10 $k \lambda$ to 300 $k \lambda$.

In addition to conventional calibrations for visibilities that include flux, gain and bandpass, we observed strong quasars (3C279, 3C454.3 and 3C84) before and after their transits to derive instrumental polarizations or the leakage term. We found that the instrumental polarization is consistent to better than 0.1\% \citep{marrone2006}. This value, together with the thermal noise, defines the uncertainty in the polarization percentage of the science targets. The basic observational parameters and the list of observed sources are given in Tables 1 and 2.

The raw visibilities were calibrated in the IDL superset MIR for flux, bandpass and time dependent
gains, and were exported to MIRIAD {\citep{sault1995}. Calibration of instrumental polarizations and additional
processing were carried out in MIRIAD. The continuum data for science targets were constructed from summing the line free spectral channels in visibilities,
and were then self calibrated to further improve the dynamic range in the image. The solutions from the
self calibration were applied to the spectral line data. The typical on-source time including all
three array configurations is approximately 6 hour per target, resulting in a typical $1 \sigma$ rms noise in the
Stokes Q and U maps of 2 mJy~beam$^{-1}$. Sources with southern declination such as NGC6334 have a higher rms noise of 5-6 mJy~beam$^{-1}$  mainly due to low elevations during the observation that resulted in higher system temperatures.

\section{Results}

\subsection{Dust Continuum and Polarized Emission}

Twenty one massive star forming regions were observed in the polarization mode with the SMA. They have masses of $10^2 - 10^3$ \msun\ within a scale of $<$ 1pc, capable of forming a cluster of stars. Strong dust polarization is detected in 14 of them. During the initial survey, two objects have no detection in dust polarization at a sensitivity of 5 mJy. The remaining targets showed marginal detection at a level of 4$\sigma$, and are followed up with additional observations to improve the signal to noise ratios in the map. This paper presents the sample of
14 molecular clumps with significant polarization. They have luminosities of $> 10^3$ \lsun, and are known to harbor massive protostars. 
Figure 1 presents the dust continuum emission overlaid by the orientation of dust polarization (E field, left panels), and the orientation of the plane-of-sky component of magnetic fields (right panels).  The length of the line segments on the left panels is  proportional to the
percentage of polarization. The color scales show the polarized and Stokes I emission.

The continuum emission (Stokes I) is detected strongly with signal-to-noise ratios greater than 50 for all the targets. The noise level in the Stokes I continuum is limited by the dynamic range due to a relatively sparse UV coverage. As a result, the noise level in these maps is set by the systematics in the dirty beam pattern and is higher than the noise in the Stokes Q and U maps. We classify the sample into elongated/filamentary and irregular at the parsec scale based on the morphology of the continuum emission obtained from the SMA, and the morphology of the cloud at a scale $> 1$ pc from single dish ground-based telescope and Herschel Space Telescope. For example, sources NGC 6334I/In/IV/V are part of filamentary cloud seen in continuum with ground-based \citep{sandell2000,munoz2007} and space-born Herschel observations \citep{russeil2013}. Likewise, sources G34.4, W51E2, W51N and DR21 (OH) are part of filamentary clouds \citep{rathborne2006,tang2010,motte2007}. In addition, the continuum emission of G240, IRAS18360 and G35.2N \citep{qiu2009a,qiu2011,qiu2013} also appear to be elongated, thus are included in the category of filamentary objects. Despite its filamentary appearance, we exclude NGC2264 C1 from the list because of its linear size of 0.05 pc, which corresponds to the size of a dense core.

The polarized emission $I_{pol}$ is computed from the intensity of Stokes Q and U according to $\sqrt{Q^2 + U^2 - \sigma_Q^2}$. Here $Q$ and $U$ and $\sigma_Q$ are flux intensity and the rms noise in Q/U images, respectively. The polarization percentage is then computed from the ratio between the polarized flux and the Stokes I flux. The polarization angle $\theta$, measured counter clockwise from the north direction to the east, is computed from tan$2\theta = U/Q$. The maps of the polarized emission, polarization percentage, and polarization angle are produced using the MIRIAD task $impol$. We used the cutoff of $5\sigma_I$ in the Stokes I maps, and $2.5\sigma_Q$  in the Stokes Q and U maps when deriving the polarization flux, polarization percentage and polarization angle. Here $\sigma_I$ is the rms noise in the Stokes I image. 
The uncertainty in polarization angles follows $28.65^\circ {\sigma_p \over p}$, where $\sigma_p$ and $p$ are noise and signal in polarization \citep{serkowski1974,naghizadeh1993}. The majority of polarization detections have signal-to-noise ratios of $> 5$, thus, their associated errors in polarization angles are $< 5.8^\circ$.

As shown in Figure 1, the distribution of the polarized emission ranges from spatially compact
in G192 to more extended in sources such as DR21 (OH). Except G192 and NGC2264 C1, spatially extended polarized emission is detected in G240, NGC 6334 I/In/IV and V, G35.2N, IRAS18360, G34.41, W51E2, W51N and DR21(OH). The polarized fluxes range from  10 to 50 mJy at a beam size of $1''.5$. The polarization percentage varies from 1\% to as high as 10\% in sources such as NGC6334In. Among the sources with spatially resolved polarized emission, DR21 (OH) shows polarization with large angular dispersions. The polarization segments are relatively ordered in sources G240, NGC6334 I/In, G34.43, G35.2N and W51E2. Sources NGC6334 IV/V, IRAS18360, W51E2 have moderate dispersions in the polarization angle.

\subsection{Molecular Outflows}

Molecular outflows often mark the birth of protostars.  During the phase of protostellar accretion, infalling gas with excess angular momenta is ejected in a wind. The detailed process involved in launching the wind is unclear for massive outflows \citep{qiu2007}. However, it is likely related to magnetic fields, similar to outflows in low-mass protostars \citep{shang2007}. The wind, which travels at several 100 \kms-1\ \citep[e.g. HH80/81][]{marti1998}, interacts with the ambient gas and produces molecular outflows seen in CO \citep{zhang2001} and other tracers \citep{qiu2007}. Therefore, the orientation of outflows defines the axis of the accretion disk that is not spatially resolved in the SMA observations.

The right panels of Figure 1 present molecular outflows identified in the CO 3-2 transition obtained simultaneously in the SMA polarization observations. The blue and red shifted CO emission are plotted in blue and red contours. For W51E2, we use the CO 3-2 outflow data from \citet{shi2010}. We also
mark the orientation of outflows with arrows. In many cases, blue and red shifted outflows can be paired in a bipolar fashion. However, there are outflows that appear to be unipolar. We identified a total of 21 outflows in our sample. At the first glance, the major axis of some outflows, e.g., G240, is parallel to the magnetic fields. Other outflows, such as those in G192 and G35.2N , have major axes perpendicular to the orientation of the magnetic field. There are also outflows that are oriented between $0^\circ$ to $90^\circ$ with respect to the magnetic field. Some outflow pairs emanate from the same cores, but with the outflow axes more than 60$^\circ$ apart (e.g., DR21(OH)).

\subsection{Comparison with Clump-Scale Polarization and Morphology}

One of the questions to be addressed in this survey is how magnetic fields at scales of $ \lsim 0.1$ pc compare with that of the parsec-scale magnetic fields revealed by the single dish telescopes. In our sample, sources with published single dish polarization measurements are NGC6334 I/In/VI/V, G34.43, G35.2N, DR21(OH), W51E2, W51N and DR21(OH) \citep{matthews2009,cortes2008,dotson2010}.  These data, measured at a spatial resolution from 10$''$ to 20$''$, correspond to the average polarization direction at a linear scale of 0.1 to 0.6 pc for source distances from 1.5 kpc to 6 kpc. We compare the polarization position angle measured with the SMA  with the position angle measured at the larger scale for the same position. The majority of the SMA data reveal polarized emission within an extent of $< 15''$. The most extended polarization emission in the SMA observations is $20''$ seen in DR21(OH). Thus, the large-scale polarization do not vary significantly across the SMA maps.

To account for the difference in linear resolution in the SMA sample due to varying source distances, we convolved the Stokes I/Q/U images to the lowest linear resolution (0.03 pc) corresponding to the source at the furthest distance. We then compute polarization angles
using the MIRIAD task {\it impol}. This normalization is needed to remove the bias toward the nearby objects which have a better linear resolution in the images, thus contribute more data points to the statistical analysis outlined below. For the nearest targets, the convolution degrades the angular resolution by a factor of 3.5. We note that the polarization angles in the convolved images resemble the maps at native SMA resolutions.

We use the convolved SMA images to compute $\mid\theta_{clump}(pol) - \theta_{SMA}(pol)\mid$ for every independent measurement in the SMA data.
Here $\theta_{clump}(pol)$ is the polarization position angle measured
​with single dish telescopes, and $\theta_{SMA}(pol)$ is the polarization position angle measured with the SMA at the scale of dense cores. For most of the targets, the spatial extent of the polarized emission is smaller than the single dish beam. For polarization sources that are more extended than the single dish beam, we interpolate the single dish data to compute the angular difference. To limit errors in polarization angles, we apply a cutoff of $\sigma > 3.5 \sigma_{Q}$ to the SMA polarization data, although the results remain the same if a lower cutoff of 3 or 2.5 $\sigma$ is applied. Since the direction of magnetic field vectors is not determined in polarization measurements, we limit the angular difference $\mid{\theta_{clump}(pol) - \theta_{SMA}(pol)}\mid$ from 0$^\circ$ to 90$^\circ$. A 0$^\circ$ difference corresponds to the case that magnetic fields in the SMA image are aligned with the large-scale magnetic fields, whereas a 90$^\circ$ difference implies that the two are perpendicular. 

Figure 2 presents the distribution of the angular difference between the polarization measurements with the single dish and those measured with the SMA. The figure shows that there is a large number of SMA polarization segments aligned within 40$^\circ$ with the large-scale polarization. Among the 173 independent polarization measurements at a scale of 0.03 pc, over 60\% of them are at position angles within 40$^\circ$ of the large-scale polarization. There is a lack of polarization segments when the angular difference lies between 40$^\circ$ to 80$^\circ$. There is another peak at an angular difference around 90$^\circ$. Overall, the angular difference of polarization segments exhibits a bimodal distribution: one group of polarization segments maintain the polarization orientation of their parental clump; the other group of polarization segments are perpendicular to the polarization orientation of their parental clump. This finding implies that magnetic fields at the core scale of 0.03 pc  are organized. Otherwise, randomly oriented fields would render an equal probability of angular distribution in Figure 2. Furthermore, Figure 2 indicates that magnetic fields at the 0.03 pc scale are correlated with the mean field orientation in their parental clump. They are either aligned within 40$^\circ$ of the pc-scale field or perpendicular to the parsec-scale field.

We also examine how polarization at the 0.03 pc scale is related to the major axis of the parsec-scale clump $\theta_{clump}(maj)$. Sources G240, NGC6334I/In/IV, G34.4, IRAS18360 and G35.2N, W51E2, W51N and DR21 (OH) are either part of a large-scale filament, or exhibit a parsec-scale elongated morphology. NGC2264 C1 is excluded from the list despite its elongated structure since its linear extent is 0.05 pc,  much smaller than a clump. However, including it in the analysis would not change the outcome since it would contribute only one data point to the statistics.
For all these elongated objects except DR21 (OH), we measure $\theta_{clump}(maj)$, the position angle of the major axis of the clump, by fitting a two dimensional elliptical Gaussian profile to the dust continuum emission from the SMA. For DR21 (OH) whose structure in the SMA map does not have a well-defined elongation (Girart et al. 2013 suggests that the clump is nearly face-one), we use the larger scale filament \citep{motte2007} to define its major axis.  We compare the position angle  of the major axis of the clump $\theta_{clump}(maj)$ with $\theta_{SMA}(pol)$. 
Figure 3 presents the distribution of angular difference, $\mid{\theta_{clump}(maj) - \theta_{SMA}(pol)}\mid$, between the filament major axis and the  polarization from the SMA. Similar to the previous case, we limit the angular difference from 0$^\circ$ to 90$^\circ$.
As shown in Figure 3, there is a large number of segments with
​their polarization parallel to the major axis of filaments. Of the 197 polarization segments, over 60\% of them are aligned with the major axis of the clump. There appears to be another peak of polarization segments with angular differences near 90$^\circ$. In between the two peaks, there is a lack of polarization segments between the angular difference from 40$^\circ$ to 80$^\circ$.  Overall, the angular difference of the segments appears in a bimodal distribution: One group of  polarization segments is aligned with the major axis of their parental clump; the other group of polarization segments is perpendicular to the major axis of their parental clump.

\subsection{Comparison with Molecular Outflows}

We further compare the outflow axis seen in the CO J = 3-2, $\theta_{outflow}$, obtained from the SMA with the position angle of magnetic fields $\theta_{SMA}(B)$ at the origin of the outflow. Here $\theta_{SMA}(B)$ is the field orientation measured within one SMA synthesized beam. When the blue and red shifted lobes form an apparent bipolar outflow, only one position angle is reported. In cases of unipolar outflows, we measure the position angle from the single outflow lobe. A total of 21 outflows is revealed in the sample. Figure 4 presents the angular difference between the outflow axis and the position angle of the magnetic fields $\mid\theta_{outflow} - \theta_{SMA}(B)\mid$. Due to a lack of information on the direction of the magnetic field, we limit the angular difference from 0$^\circ$ to 90$^\circ$. A 0$^\circ$ difference corresponds to the outflow axis being in parallel to the major axis of the outflow. Because of the small number of outflows in the statistics, we refrain from interpreting the detailed structure in the distribution. Overall, there appears to be no strong correlation between the outflow axis and the magnetic field orientation in the core from which outflows originate.​

\section{Discussion}

\subsection{Effect of Geometric Projection and Statistical Analysis}

Before we interpret the results in Figures 2, 3 and 4, we investigate the projection effect of two vectors in a three-dimensional space intersecting at an angle $\alpha$. We pose the question that when the two vectors 
are randomly oriented in space, what is its projected angle $\beta$ onto a plane? This question is relevant to the 
study here since the observations only provide the polarization component projected to the plane of the sky without 
information along the line of sight \citep[see also][]{tassis2009, hull2013}. For the majority of the vectors oriented within the 2$\pi$ solid angle, one 
expects that the projected angle $\beta$ will be smaller than $\alpha$. However, certain orientations of vectors can 
render  $\beta$ greater than $\alpha$. For example, if two vectors intersect at the origin of an XYZ Cartesian coordinate 
system, with one line being in the XZ plane, and the other line lying in the YZ plane, the angle $\alpha$ can vary 
from 0$^\circ$ to 90$^\circ$, while the projected angle $\beta$ on to the XY plane is always 90$^\circ$.

In order to simulate the projection effect, we consider a pair of randomly oriented vectors uniformly distributed in a three-dimensional space. We select all pairs with intrinsic angles $\alpha$ from 0$^\circ$ to 40$^\circ$ and from 80$^\circ$ to 90$^\circ$. The probability distributions of the projected angle $\beta$ are plotted in Figures 2 and 3. As expected, small projected angles imply a high probability of 
small intrinsic angles $\alpha$. Likewise, large projected angles close to $90^\circ$ imply a high probability of 
large intrinsic angles. A randomly oriented vector pair in a $4 \pi$ solid angle gives an equal probability distribution of 
projected angles $\beta$ from 0$^\circ$ to 90$^\circ$.

The simulated distributions of angles described above are compared with 
the observed angular difference between the clump-scale and 
the core-scale magnetic field orientations shown in Figure 2. The simulated results are shown in Figure 2 as dashed and dotted 
lines for angles from 0$^\circ$ to 40$^\circ$ and from 80$^\circ$ to 90$^\circ$, respectively. We run the 
Kolmogorov-Smirnov (K-S) test to determine the probability that the two distributions are drawn from the same 
parent population. When comparing the observed distribution with the simulated distribution of angles 
of 0$^\circ$ to 40$^\circ$, the probability that the two groups are drawn from the same population 
is $9.3 \times 10^{-4}$. The probability that the observed distribution and the simulated distribution of 
angles from 80$^\circ$ to 90$^\circ$ are drawn from the same parent population is $4.3 \times 10^{-9}$. 
Similarly, the probability that the observed distribution and a flat distribution of angles 
from 0$^\circ$ to 90$^\circ$ are drawn from the same parent population is $2.4 \times 10^{-3}$. On the other hand, 
if we combine the two simulated groups of distributions, e.g.,  from 0$^\circ$ to 40$^\circ$ and from 80$^\circ$ to 90$^\circ$, with a ratio of $5 : 3$ (shown as dashed-dotted line in 
Figure 2), then we find that the probability of the observed distribution and the combined distribution are drawn 
from the same parent population is 0.93. 

In addition to the K-S test, we also use circular statistics to test the hypothesis that the observed distribution and the simulated distributions are drawn from the same distribution. Since data presented in Figure 2 deal with directions, circular statistics are more appropriate for the analysis. We performed Watson test for two sample directional data \citep{Jammalamadaka2001}. When comparing the observed distribution and a simulated distribution of angles
from 0$^\circ$ to 40$^\circ$, from 80$^\circ$ to 90$^\circ$, and a random distribution, we found a probability of
0.001 to 0.01, $< 0.001$, and $< 0.001$, respectively. Both the K-S and Watson test reject the hypothesis that the observed distribution of angles are randomly oriented.
It appears that the observed distributions likely comprise
of two populations of angles: one group with intrinsic angles ($\mid{\theta_{clump}(pol) - \theta_{SMA}(pol)}\mid$)
from 0$^\circ$ to 40$^\circ$, and the other group with intrinsic angles ($\mid{\theta_{clump}(pol) - \theta_{SMA}(pol)}\mid$)
distributed from 80$^\circ$ to 90$^\circ$.

We also run the K-S test on the distribution of the angle difference $\mid\theta_{clump}(maj) - \theta_{SMA}(pol)\mid$ as shown in Figure 3. The simulated results are shown as dashed and dotted 
lines for angles from 0$^\circ$ to 40$^\circ$ and from 80$^\circ$ to 90$^\circ$, respectively. We found that probabilities that the observed distribution and simulated distribution of angles from 0$^\circ$ to 40$^\circ$, from 80$^\circ$ to 90$^\circ$, and a flat distribution are drawn from the same parent population are $3.4 \times 10^{-3}$,  $3.8 \times 10^{-10}$, and  $6.4 \times 10^{-3}$, respectively. If one combines the two simulated distributions of angles from 0$^\circ$ to 40$^\circ$ and from 80$^\circ$ to 90$^\circ$ at a ratio of $5 : 3$, the probability that the observed distribution and the combined distribution is drawn from the same population is 0.96. The results of the K-S tests are summarized in Table 3.

The comparison between the outflow major axis and the orientation of magnetic fields in the core is limited by small number of statistics.
Nevertheless, there appears to be a larger number of outflows (a total of 9 out of 21) with axes within 70$^\circ$ to 90$^\circ$ of the magnetic field orientation in the core. In addition, there is another peak in the distribution close to 0$^\circ$. However, that peak consists of only 5 outflows with their major axes within 10$^\circ$ of the magnetic field. Overall, the distribution in Figure 4 is consistent with a scenario that outflow orientation does not correlate with the orientation of magnetic fields in the core. 

\subsection{The Role of Magnetic Fields in the Formation of Massive Cores}

Massive stars form predominantly in the cluster environment in association with a number of lower mass objects. 
The basic unit in which massive stars form is a parsec-scale molecular clump with typical masses of $10^2 - 10^3$ \msun. 
The collapse and fragmentation of massive clumps give rise to cores with sizes of $\lsim$ 0.1 pc at higher densities. 
One of the outstanding questions 
involved in this process is the role of magnetic fields during the dynamical evolution of molecular clumps and dense 
cores. The SMA observations reveal polarization at a linear scale of 0.03 pc which corresponds to 
scales of dense molecular cores hosting individual or multiple protostars. Therefore, the SMA data unveil the plane-of-sky magnetic field information at dense core scales. Given these considerations, the results in Figure 2 demonstrate (1) magnetic fields on dense core scales in a cluster forming environment are
not randomly distributed, but are organized. Otherwise, one would expect a flat distribution instead of a
bimodal distribution; (2) Magnetic fields on core scales tend to be parallel or perpendicular to the field orientations in their parental clumps.

Numerical simulations of magnetized molecular clouds with turbulence offer guidance on deciphering the dynamic role of magnetic fields based on the field orientation. When magnetic field is strong, e.g., the ratio of gas pressure to magnetic pressure $\beta = {P_{th} \over P_B} \ll 1$, the field is less disturbed and appears to be organized. In this regime, dense gas filaments form with their long axes perpendicular to the magnetic field, while the lower density gas in the cloud forms striations connected to the dense filament, which are aligned parallel to the magnetic fields \citep{ostriker2001, nakamura2008, vanloo2014}. On the contrary, the field appears to be more randomly oriented in the weak field regime ($\beta \gg 1$). In light of the theoretical and numerical work,
the fact that the majority of the dense cores have their magnetic fields more or less aligned with or perpendicular to those in the 
clump scale indicates that the magnetic field plays a dynamically important role in core formation. The field is strong enough on the clump scale to guide 
the concentration of clump material along the field lines into dense cores. When cores become massive enough, 
gravitational contraction may become important. The contraction is expected to  pinch the magnetic field inside the 
core, but does not necessarily lead to a large misalignment between the averaged magnetic field on the core scale 
(probed by our SMA observations) and that on the clump scale (probed by single dish observations). Turbulence can 
in principle distort the magnetic fields on both the clump and core scale. In our sample, it does not appear to be 
strong enough to randomize the field directions, as would be the case if the field were dynamically 
insignificant.

The findings in Figure 2 are further corroborated by the comparison of magnetic fields in the core scale and the major axis of 
their parental clumps. In the presence of a strong magnetic field, molecular gas is expected to collapse along 
the field lines to form filamentary structures either parallel or perpendicular to the field \citep{ostriker2001, nakamura2008, vanloo2014}. Polarimetric observations in optical, 
infrared and radio wavelengths also found that the relative alignment of the large-scale magnetic fields and the major axis of filaments are in a bimodal distribution \citep{li2013} similar to that in Figure 3. For the SMA sample, those with large-scale (sub)mm polarimetric 
measurements, $i. e.$, NGC6334, G34.4, G35.2N, W51E2, W51N and DR21(OH) \citep{matthews2009,cortes2008,dotson2010}, show that the magnetic field is roughly 
perpendicular to the major axis of the filaments. If the major axis of the parsec-scale clumps traces the large-scale magnetic fields, the fact that the majority of dense cores  have magnetic fields perpendicular to the clumps 
as exhibited in Figure 3 serves as another indication that magnetic fields play an important role in the dynamic 
evolution of molecular clumps.

The apparent bimodal distributions seen in Figures 2 and 3 are remarkable. A distribution similar to that in Figure 3 is expected in a regime of strong magnetic fields. As suggested in simulations \citep{nagai1998,nakamura2008,li2013,vanloo2014}, the segments perpendicular to the clump-scale field could arise from lower density gas. We examined the densities derived from dust emission against $\mid\theta_{clump}(pol) - \theta_{SMA}(pol)\mid$ as well as $\mid\theta_{clump}(maj) - \theta_{SMA}(pol)\mid$,
 and found no dominant trends among the two groups. This could be, in part, due to the fact that the projection of angles discussed in Section 4.1 mixes the two groups together. Therefore, both high density and low density cores can have the same angular difference $\theta_{clump}(pol) - \theta_{SMA}(pol)$ due to the projection effect.  In addition, the sensitivity in the polarization data limits detections of field lines associated with the lower density gas. Polarization studies with more sensitive interferometers such as ALMA will provide insight on this issue.

\subsection{The Role of Magnetic Field at Scales of Accretion Disks}

Protostellar outflows observed in CO mark the ambient gas accelerated by the wind ejected during the accretion and 
mass assembly of a protostar. Outflows from massive protostars have been the subject of intense observational 
studies with both single dish telescopes \citep{zhang2001,beuther2002b,zhang2005a} and  
interferometers \citep[e.g.][]{cesaroni1999a, beuther2002d, sollins2004b,su2004,su2007,qiu2007,qiu2008,qiu2009a,qiu2009b,qiu2011,wang2012,duarte-cabral2013}. 
These molecular outflows are often one to two orders of magnitude more massive and  energetic in mass, 
momentum and energy than outflows from lower mass protostars \citep{zhang2005a}. While the launch process is 
still unknown, these energetic outflows are believed to be driven by a magneto hydrodynamic wind  in the disk 
perhaps analogous to the disk-wind or X-wind \citep{shang2007} that operates in low-mass protostars. Regardless of 
the driving process, molecular outflows offer insights into the accretion process that occurs at a scale 
of $\lsim$ 1000AU within a dense core. While massive ($10^2$ \msun) rotating structures, known as toroids  have 
been reported widely in the literature \citep{zhang2005b,cesaroni2007}, they represent the envelope of dense 
cores \citep{keto2010} that likely further fragment \citep{wang2011,wang2014}. In some cases where a massive protostar 
is relatively isolated, a rotationally supported accretion disk may have a diameter as large as 6000 AU, as 
reported by \citet{keto2010} in a radiative transfer modeling of multiple spectral line data in a massive 
protostar IRAS20126+4104. However, in a highly clustered environment where massive stars are predominantly formed, 
the size of an accretion disk can be much smaller due to dynamical interactions. Therefore, the axis of an outflow 
provides an indirect measure of disk orientation, which normally lies in a perpendicular direction.

The comparison between the outflow axis and the position angle of the magnetic field in the core from which the 
outflow is launched suggests that there is no preferred alignment between the two axes. This finding implies that  
magnetic fields in the accretion disk do not maintain the field direction of the core. For example,
in the DR21(OH) region, two outflows appear to originate from the same 
core, but have position angles as far apart as 60$^\circ$. Our results indicate that disk orientation is not controlled
by the magnetic field that we probe with the SMA on the core scale. This is in contrast to the simplest expectation
where the component of the angular momentum perpendicular to the magnetic field is removed more efficiently by
magnetic braking than the parallel component, which would tend to drive the angular momentum vector (and hence
the outflow direction) to align with the field direction. There are several possible explanations to the misalignment. (1) The misalignment may
indicate that the field on the disk formation scale is not strong enough to brake the angular momentum/rotation perpendicular
to the field, or the field and gas are decoupled in the dense region so that magnetic braking is weakened \citep[e.g.][]{mellon2009,padovani2013}; (2) The disk orientation is controlled by some other dynamical processes, such as angular momentum redistribution and gravitational
interactions in binary or multiple systems. In any case, the dynamical  importance of the magnetic field appears to
weaken from the core to the disk scale. 

Recently, a polarization survey of low-mass protostellar cores with CARMA \citep{hull2013} found that outflow axes do not correlate with the magnetic field orientation in the dense envelope surrounding accretion disks. This result appears to contradict with a similar 
survey carried out with the SHARP polarimeter on CSO by \citet{chapman2013} who reported a good alignment 
between protostellar outflows and magnetic field orientation in the core. As proposed by \citet{li2013}, these two results can be reconciled in that at the scale probed by CARMA, magnetic fields are dragged and twisted by rotation in disks and envelopes. Another difference is that the majority of the sources in Chapman et al.'s (small) sample have outflows close to the plane of the sky. This minimizes the projection effect in the magnetic field orientation which could be considerable in Hull et al.'s sample. However, as \citet{hull2013} pointed out, the projection effect may not explain the difference since it becomes significant only for outflows that are nearly along the line of sight, which are rare.

Although our result appears to be in agreement with \citet{hull2013}, the interpretation may not be the same. Since our sample is at a distance an order of magnitude larger than the Hull et al.'s sample, rotation in general does not produce appreciable effect on magnetic fields over the linear scale that SMA probes, except for nearby targets such as G192 \citep{liu2013}. For our sample, it is likely that dynamic interactions in binaries 
or close multiples change the disk and thus outflow orientation relative to the magnetic direction.

\subsection{Effect of Spatial Filtering}

One of the key findings in this paper (e.g. Figure 2) is derived from the comparison of polarization measurements between the single dish telescopes and the SMA. Figure 2 reveals that magnetic fields at core scales tend to maintain the mean field orientation measured by the single dish telescopes. Can the SMA data be dominated by the smooth structure seen in single dish telescopes, hence, create a false alignment between the clump-scale and the core-scale magnetic fields? The answer is negative.

While polarization angles measured by a single dish telescope represent the mean orientation within the telescope beam,
the SMA observations, like any interferometers, probe the spatially compact structures within, while filtering out extended emission.
The SMA images reveal spatial structures from 20$''$ down to $1''$, the synthesized beam of the observations. The typical densities in dense cores inferred from the Stokes I dust continuum are $>10^5$ cm$^{-3}$, at least one order of magnitude higher than the average density of their molecular clumps. The combined effect of high resolution and high density ensures that the SMA polarization data probe the gas at $\lsim$ 0.1 pc scales closely associated with star formation. One clear example is DR 21(OH). The SMA observations at high angular resolution reveal disordered polarization structure oriented in the east-west and north-south directions \citep{girart2013}. In contrast, the polarization image obtained at the same frequency band from the JCMT and CSO exhibits east-west magnetic field lines \citep{vallee2006,matthews2009}. We test the possibility that an ordered field from single dish observations is a result of either spatial filtering or spatial smoothing, the later of
which tends to render a  more uniform field orientation. We tapered the SMA images to 20$''$ and reconstructed the polarization map. Figure 5 presents a comparison of the JCMT SCUBA map with the convolved SMA image to 20$''$. As one can see, the convolved SMA image remains similar to the higher resolution SMA maps. Therefore, the single dish data probes the larger scale magnetic field, whereas interferometer maps presented here probe the underlying structures at smaller spatial scales.

\subsection{A Schematic Picture on the Role of Magnetic Fields and Limitation of the Study}

The polarization study of a large sample of massive star forming regions reveals important clues to the role of magnetic fields during the collapse and fragmentation of massive molecular clumps. Figure 6 presents a schematic picture summarizing the key findings in this paper. At the clump scale, gas at densities of $>  10^4$ \CM2\ is threaded by magnetic fields typically probed by single dish telescopes. The parsec-scale clump can be elongated and/or be part of a larger scale filament, with the major axis perpendicular to the magnetic field. The collapse of the clump leads to the formation of cores at higher densities. The magnetic field in the higher density gas appears to be either aligned with  or perpendicular to the clump-scale field.   As the gas in dense cores continues to condense to form accretion disks that drive molecular outflows, the major axis of the disk does not appear to be aligned with the magnetic field in the parental core.

While this study represents the largest sample of massive star forming regions imaged in polarization with a (sub)mm interferometer thus far, the sample included in the statistical analysis contains only ten massive molecular clumps. Therefore, results could be biased by the collection of targets that by chance have organized magnetic fields at core scales. Among most of the sources imaged, the detected polarized emission extends to a fraction of the Stokes I image. The statistical analysis does not take into account these cores with no polarization detections. Limited studies suggest that polarization can be distorted by stellar feedback. For example, radiation and ionization from HII regions can influence magnetic field orientation \citep{tang2009a}.  The sample in this paper is mostly at an earlier stage of evolution prior to the HII region phase. Therefore, the effect of feedback from an HII region is minimal in our sample. At the sensitivity of the SMA, detecting dust polarization in massive IRDC (infrared dark cloud) clumps is still difficult due to the faintness of the sources. However, these objects may contain the pristine physical conditions of massive star formation with little influence from stellar feedback. Measuring polarization in these objects will be possible with more sensitive telescopes such as ALMA.

While the results presented here are limited by the sample size, it is apparent that the technique of comparing magnetic fields at different spatial scales shows a great promise in constraining the relative role of magnetic fields versus turbulence and gravity/rotation. Such a technique can be applied when polarization measurements of a much larger sample become feasible. Interferometers such as SMA and CARMA that are currently equipped with polarimeters are on the verge of their capability to undertake large observing programs. The scientific return of such programs is apparent as demonstrated in the progress made in studying magnetic fields in low and high-mass star formation \citep[e.g.][and this work]{hull2013}. While these instruments will continue to play a role in polarimetric studies, they pave the way for further studies of much larger samples with ALMA as its polarimetric capability continues to mature. It is hopeful that more statistically robust studies based on much larger samples will come to fruition in the next decade.

\section {Conclusion}

We present SMA observations of a sample of 14 massive molecular clumps in the 345 GHz band (870 $\mu$m) to investigate the role of magnetic fields in massive star formation. The key findings are summarized below:

\begin{enumerate}

\item The spatially resolved polarization images show that magnetic fields at the core scale of 0.01 to 0.1 pc tend to be more organized rather than chaotic;

\item By comparing magnetic field orientations at a spatial scale of 0.03 pc (approximately the spatial scale of dense molecular cores) with the parsec-scale mean field orientations from single dish telescopes, we found a bimodal distribution in the alignment of the two axes: Among the cores with polarization detection, over 60\% have their fields aligned to within 40$^\circ$ of the parsec-scale mean magnetic field orientation; while the remaining group of cores have their fields perpendicular ($80^\circ - 90^\circ$) to the  mean magnetic field orientation. From a K-S test, the probability that the observed distribution and the distribution from randomly aligned axes are drawn from the same population is $2.4 \times 10^{-3}$;

\item  Similarly, by comparing magnetic field orientations at core scales with the major axis of molecular clumps, we found that over 60\% of cores have their fields perpendicular to the major axis;

\item  The major axes of molecular outflows do not appear to be correlated with the orientation of magnetic fields in cores from which outflows are launched;

\item  These findings demonstrate that  magnetic fields play an important role during the collapse and fragmentation of parsec-scale clumps, and the formation of dense cores. The magnetic field is strong enough on the clump scale to guide the concentration of clump material along the field lines into dense cores. Furthermore, no alignment between the outflow axes and the field orientation in the core indicates either that the field on the disk formation scale is not strong enough to brake the rotation perpendicular to the field, or that the disk orientation is controlled by some other dynamical processes, such as gravitational interactions in binary or multiple systems.

\end{enumerate}

\acknowledgements
We are indebted to the SMA staff for their effort in instrument development and observing support. Without their dedication,  this large polarization study would not have been possible. Q. Z. is grateful to F. H. Shu for his encouragement, and R. L. Plambeck and C. L. H. Hull for enlightening discussions. J.M.G. and C.J. are supported by the Spanish MINECO AYA2011-30228-C03 and Catalan AGAUR 2009SGR1172 grants. S.P.L and T.C.C. are supported by the Ministry of Science and Technology Taiwan with grants NSC 101-2119-M-007-004 and MOST
102-2119-M-007-004-MY3.

\begin{table*}
\caption{List of Observational Parameters}
\label{tab:1}
\begin{center}
\begin{tabular}{l|ccccc}
\hline \hline
 Date of      & Configuration  &  Bandwidth &     $\tau$   & Number of  & Polarization \\
 Observations &             &     (GHz)     & at 225 GHz   & Antennas   & Calibrators   \\
\hline
 2008/05/07   & Compact     &     2   &  0.1  & 6  & 3C273      \\ 
 2008/05/25   & Compact     &     2   &  0.1  & 8  & 3C273      \\ 
 2009/05/15   & Compact     &     2   &  0.1  & 7  & 3C273      \\ 
 2010/05/06   & Compact     &     4   &  0.05  & 7  & 3C273/3C454.3 \\ 
 2011/06/18   & Compact     &     4   &  0.06-0.1  & 7  & 3C279/3C454.3      \\
 2011/06/19   & Compact     &     4   &  0.06   & 7  & 3C279/3C454.3      \\
 2011/06/21   & Compact     &     4   &  0.06   & 7  & 3C279/3C454.3      \\
 2011/07/02   & Subcompact     &     4   &  0.09   & 7  & 3C279/3C454.3      \\
 2011/07/05   & Subcompact     &     4   &  0.1   & 7  & 3C279/3C454.3      \\
 2011/07/05   & Subcompact     &     4   &  0.04   & 7  & 3C279/3C454.3      \\
 2011/07/13   & Subcompact     &     4   &  0.05-0.1  & 7  & 3C279/3C454.3      \\
 2011/07/20   & Extended     &     4   &  0.05  & 8  & 3C279/3C454.3      \\
 2011/07/21   & Extended     &     4   &  0.05  & 8  & 3C279/3C454.3      \\
 2011/07/23   & Extended     &     4   &  0.05  & 8  & 3C279/3C454.3      \\
 2011/10/17   & Compact     &     4   &  0.04   & 7  & 3C279/3C454.3      \\
 2011/11/29   & Compact     &     4   &  0.05-0.15   & 8  & 3C84      \\
 2011/12/10   & Compact     &     4   &  0.05   & 8  & 3C84      \\
 2012/01/06   & Subcompact  &     4   &  0.05   & 6 & 3C84      \\
 2012/01/10   & Subcompact  &     4   &  0.1   & 7 & 3C84      \\
 2012/02/02   & Extended  &     4   &  0.04   & 7 & 3C84      \\
 2012/03/27   & Extended  &     4   &  0.1   & 7 & 3C84      \\
 2012/03/28   & Extended  &     4   &  0.07   & 6 & 3C84      \\
\hline
\end{tabular}
\end{center}
\end{table*}

\newpage
\begin{table*}
\caption{Sample of Massive Star Forming Regions}
\label{tab:1}
\begin{center}
\begin{tabular}{l|r|r|c|c|l}
\hline
Source & $\alpha$ (J2000) & $\delta$ (J2000) & $V_{\rm lsr}$  & D  & Date of Observations/Config$^a$ \\
       &                  &                  &  (km s$^{-1}$) &     (kpc) &                         \\
\hline
G192       & 05:58:13.55 & 16:31:58.3 & 5.7 & 1.52 & 111210/C, 120126/E, 120202/E \\
NGC2264C1 & 06:41:17.95 & 09:29:03.0 & 7   & 0.3 & 111129/C, 120110/S, 120126/E, 120202/E  \\

G240.31 & 07:44:51.97 & -24:07:42.5   & 67.5 & 4.7 & 111017/C, 111210/C,120126/E, 120202/E \\
        &             &               &      &     & 120327/E\\
NGC6334VI      & 17:20:19.10 & -35:54:45.0 & -8 & 1.7 & 090515/C, 110705/S, 110720/E \\
NGC6334I      & 17:20:53.44 & -35:47:02.2 & -8 &  1.7 & 080525/C, 110712/S, 110721/E \\
NGC6334In    & 17:20:54.63 & -35:45:08.5 & -8 &  1.7 & 080507/C, 110709/S, 110720/E \\
NGC6334V    & 17:19:57.40 & -35:57:46.0  & -8 &  1.7 &110618/C, 110702/S, 110721/E  \\

G34.4.0 & 18:53:18.01 & 01:25:25.6 & 57   & 1.57 & 110619/C, 110705/S, 110720/E, 110721/E \\
        &             &            &      &      & 110723/E \\
G34.4.1 & 18:53:18.68 & 01:24:47.2 & 57 & 1.57 & 110619/C, 110705/S, 110720/E, 110723/E \\

G35.2N & 18:58:12.96 & 01:40:36.9 & 34 & 2.19 & 100506/C, 110702/S, 110721/E, 110723/E \\
       &             &            &    &      & 120328/E\\
I18360 & 18:38:40.74 &  -05:35:04.2 & 103.5 & 4.8 & 100506/C, 100918/E, 110709/S \\

W51E2/E8$^{b}$ & 19:23:43.95 & 14:30:34.0 & 58 & 6.3 & 110621/C  \\

W51N$^{b}$ & 19:23:40.05 & 14:31:05.0 & 58 & 5.4 & 110621/C, 110712/S \\

DR21(OH)    & 20:39:01.20  & 42:22:48.5 & -3.5 & 1.7 & 110621/C, 110720/E, 110721/E, 110723/E \\ 
            &              &            &      &     & 110718/E, 110903/V, 111017/C \\

\hline

\end{tabular}\\
\tablenotetext{a}{Date of Observations/Config. is listed in a format of $yymmdd/Configuration$, 
where configuration $S$ stands for Subcompact, $C$ for Compact, and $E$ for Extended configuration.}
\tablenotetext{b}{Maps presented in this paper also include observations in Tang et al. (2009, 2013).}
\end{center}
\end{table*}

\begin{table*}
\caption{Comparison of angle distributions between observations and simulations$^{a}$}
\label{tab:3}
\begin{center}
\begin{tabular}{l|ccccc}
\hline \hline
          & $0^\circ - 40^\circ$   &  $80^\circ -90^\circ$  &    Random   & Combined$^{b}$ \\
\hline
 Case 1$^{c}$   &  $9.3 \times 10^{-4}$ &   $4.3\times 10^{-9}$  &  $2.4\times 10^{-3}$   &  0.93    \\
 Case 2$^{c}$   &  $3.4 \times 10^{-3}$ &   $3.8\times 10^{-10}$  &  $6.4\times 10^{-3}$  & 0.96       \\
\hline
\end{tabular}
\tablenotetext{a}{The table lists probabilities that the observed distribution and simulated distributions are derived from the same parent population based on the K-S test.}
\tablenotetext{b}{Combine the distributions of $0^\circ - 40^\circ$ and $80^\circ -90^\circ$ with a ratio of $5 : 3$.}
\tablenotetext{c}{Cases 1 and 2 correspond to the comparison of the observed distribution and the simulated distributions in Figures 2 and 3, respectively.}
\end{center}
\end{table*}

\pagebreak
Fig. 1:  {\bf Left:} Continuum emission at 345 GHz (870 $\mu$m) obtained with the SMA. The black contours mark the continuum emission in Stokes I. The color scales in units of mJy/beam represent the polarized emission $\sqrt{Q^2 + U^2 - \sigma^2}$. Line segments in red denote the polarization (E field). The length of the segments is proportional to the percentage of the polarization, which is marked at the upper-left corner of the panel. The linear length scale of the map is plotted at the bottom-right of the panel. The synthesized beam of the images are plotted as the shaded ellipse at the lower-left corner of the panel.
{\bf Right:} CO outflows overlaid on the magnetic field orientations. The grey scales in unit of Jy/beam represent the Stokes I emission. Line segments represent the orientation of the plane-of-sky component of magnetic fields. The blue and red contours represent the blue and red shifted outflows, which are marked by arrows. For sources that clump-scale polarization data are available, the magnetic field orientation is marked at the upper-left corner.

\pagebreak
\begin{figure}[h]
\figurenum{1a}
\resizebox{!}{12cm}{\includegraphics{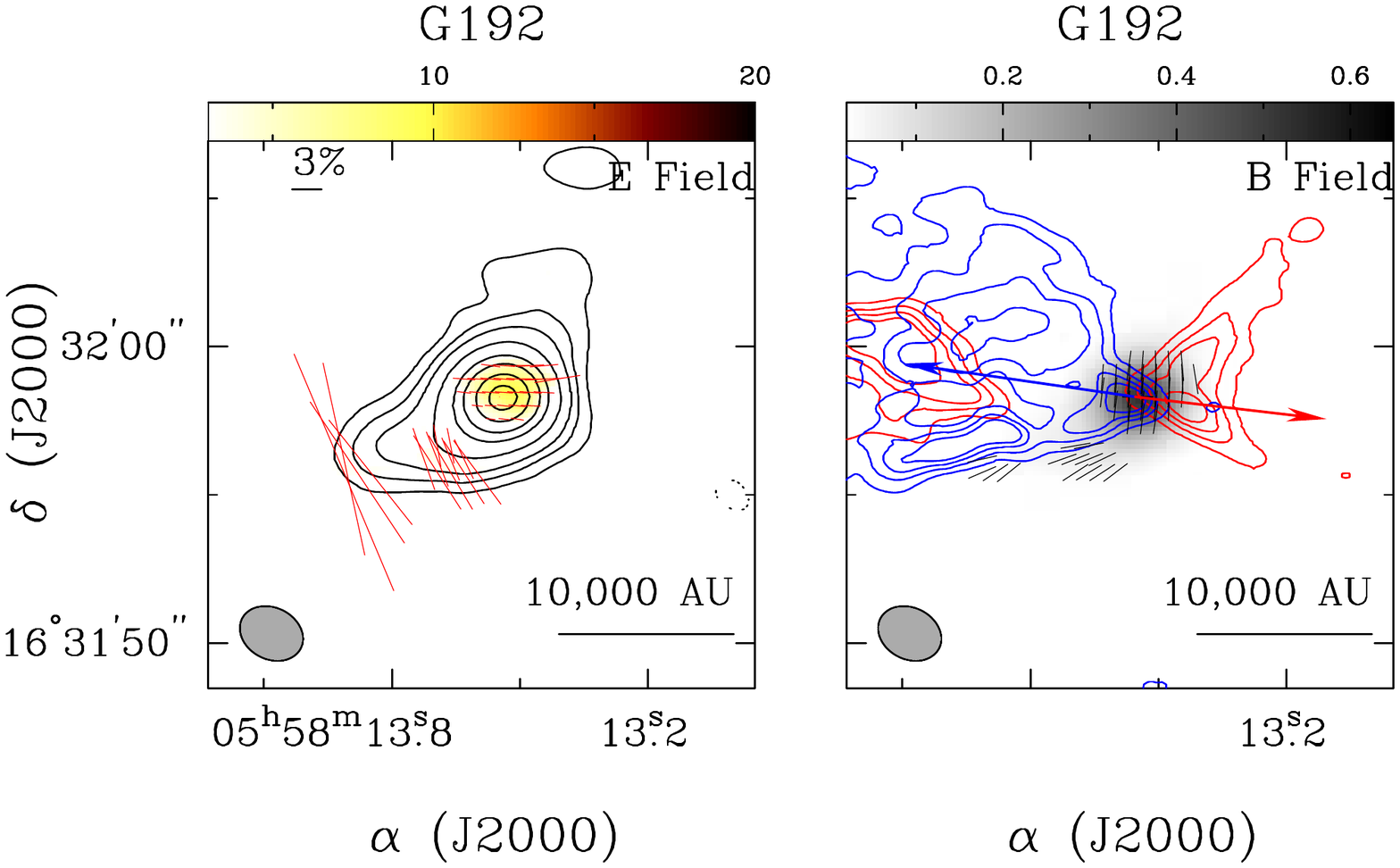}}
\caption{G192: The contour levels for the Stokes I image are $0.004 \times (\pm3, \pm6, 10, 20, 40, 60, 90, 120, 150)$ Jy~beam$^{-1}$, with a synthesized beam size of $2''.22 \times 1''.74$ at a positional angle of $-86^\circ$.}
\end{figure}

\pagebreak
\begin{figure}[h]
\figurenum{1b}
\resizebox{!}{12cm}{\includegraphics{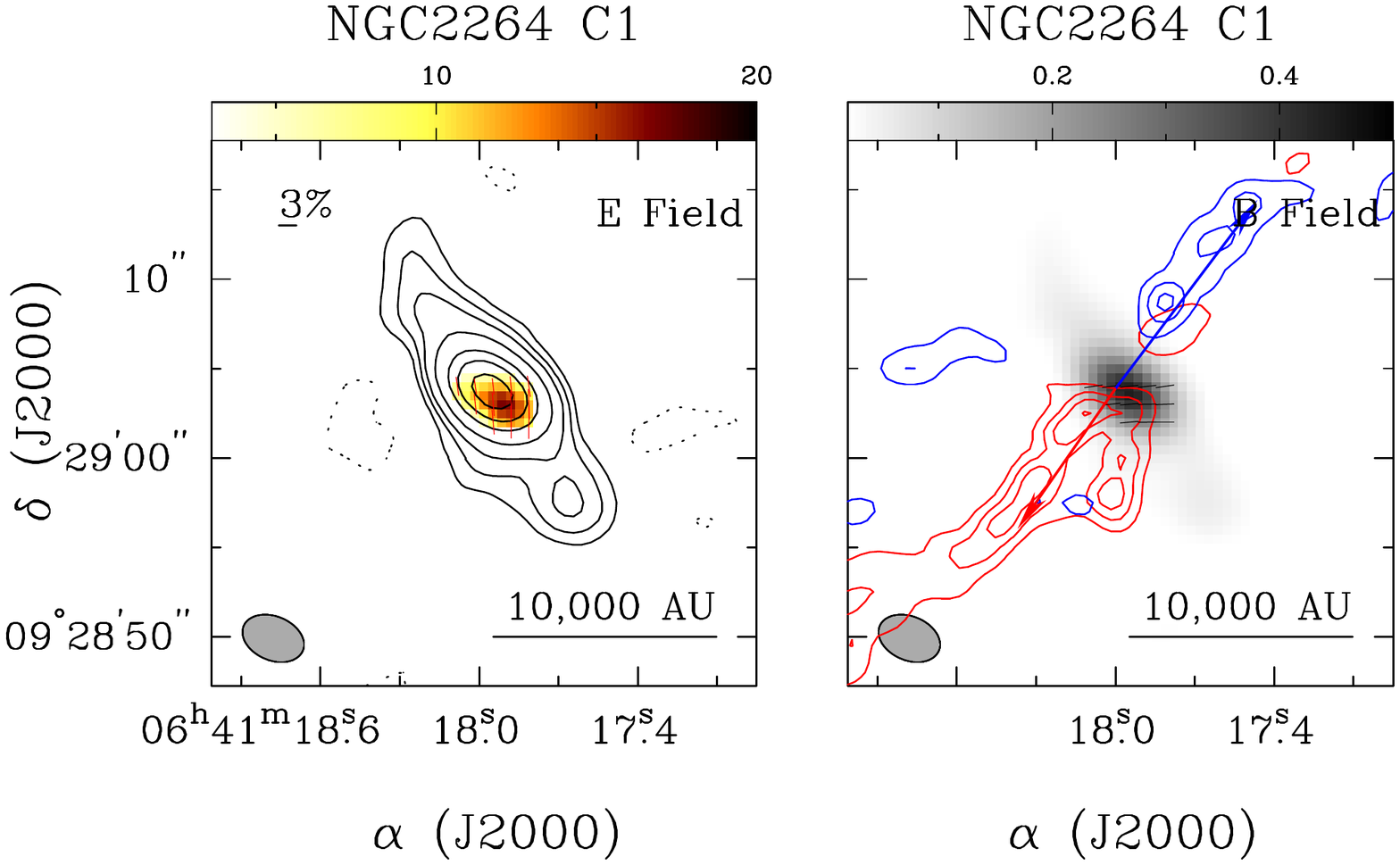}}
\caption{N2264C1: The contour levels for the Stokes I image are $0.004 \times (\pm3, \pm6, 10, 20, 40, 60, 90, 120, 150)$ Jy~beam$^{-1}$, with a synthesized beam size of $3''.59 \times 2''.46$ at a positional angle of $67^\circ$.}
\end{figure}

\pagebreak
\begin{figure}[h]
\figurenum{1c}
\resizebox{!}{12cm}{\includegraphics{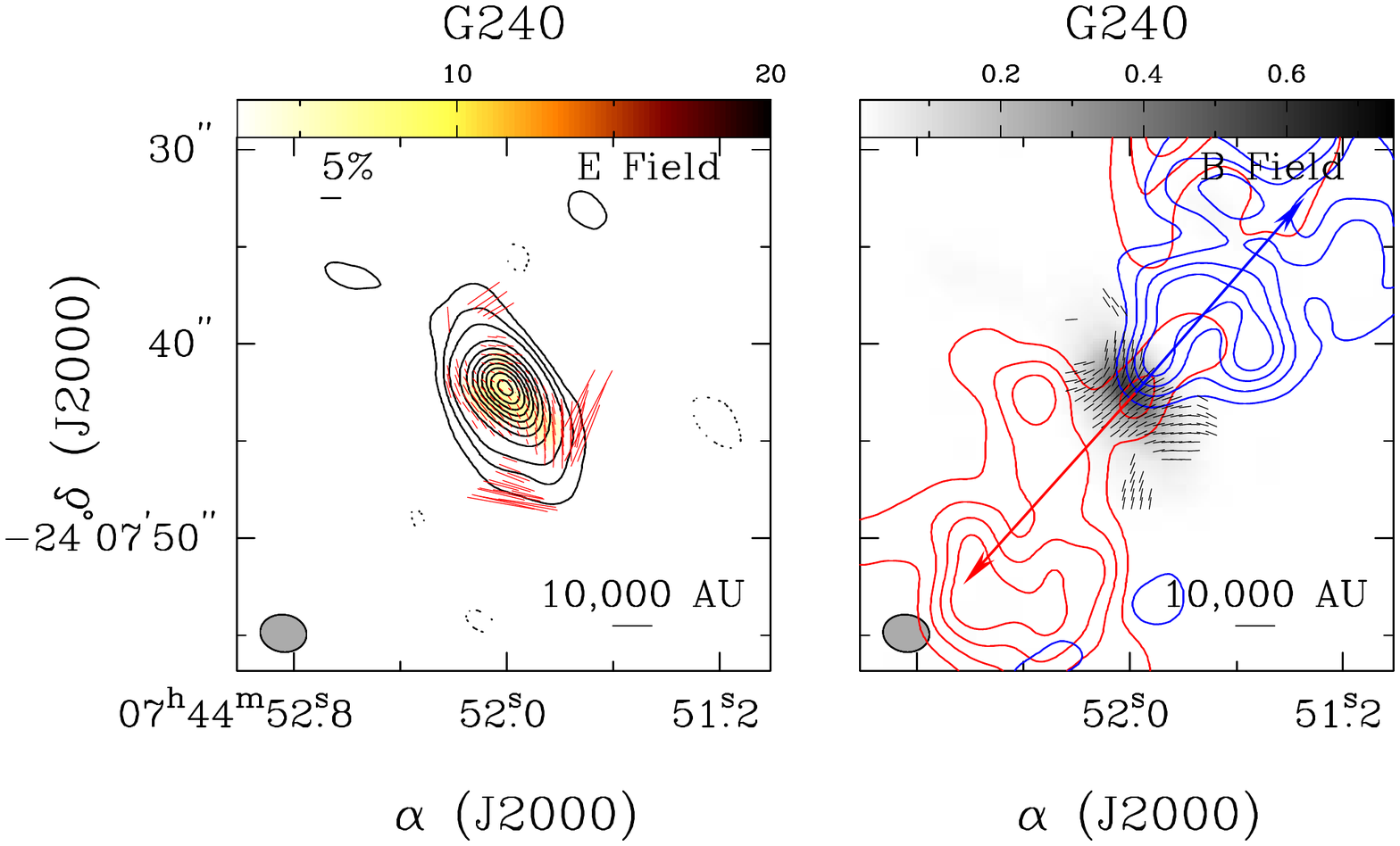}}
\caption{G240: The contour levels for the Stokes I image are $0.006 \times (\pm3, \pm6, 10, 20, 30, 40, 50, 60, 70, 80, 90, 100, 110, 120)$ Jy~beam$^{-1}$, with a synthesized beam size of $2''.39 \times 1''.92$ at a positional angle of $28^\circ$.}
\end{figure}

\pagebreak
\begin{figure}[h]
\figurenum{1d}
\resizebox{!}{12cm}{\includegraphics{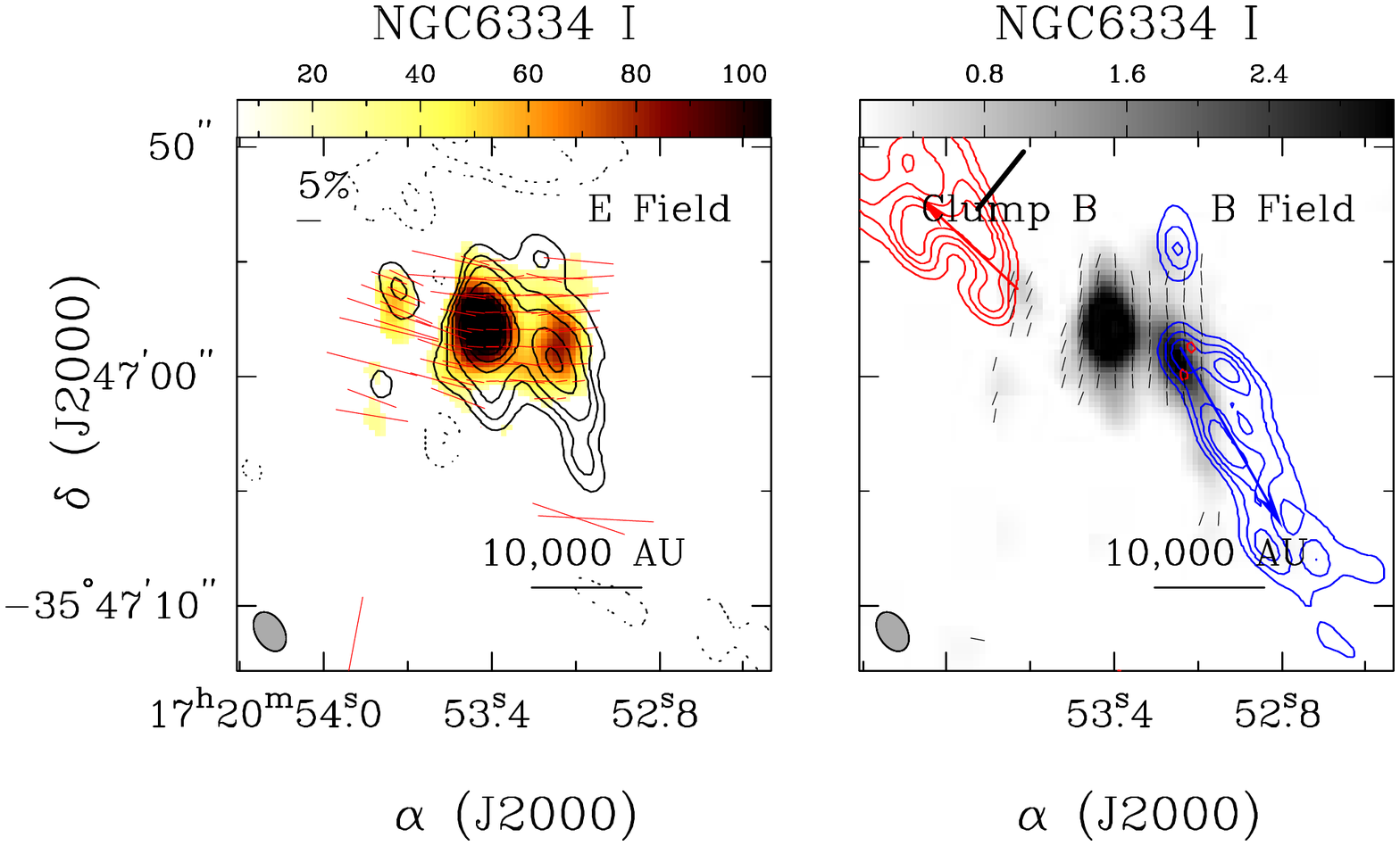}}
\caption{NGC6334I: The contour levels for the Stokes I image are $0.1 \times (\pm3, \pm6, 10, 20, 30, 40, 50, 60, 70, 80, 90, 100, 110, 120)$ Jy~beam$^{-1}$, with a synthesized beam size of $1''.85 \times 1''.23$ at a positional angle of $31^\circ$. }
\end{figure}

\pagebreak
\begin{figure}[h]
\figurenum{1e}
\resizebox{!}{12cm}{\includegraphics{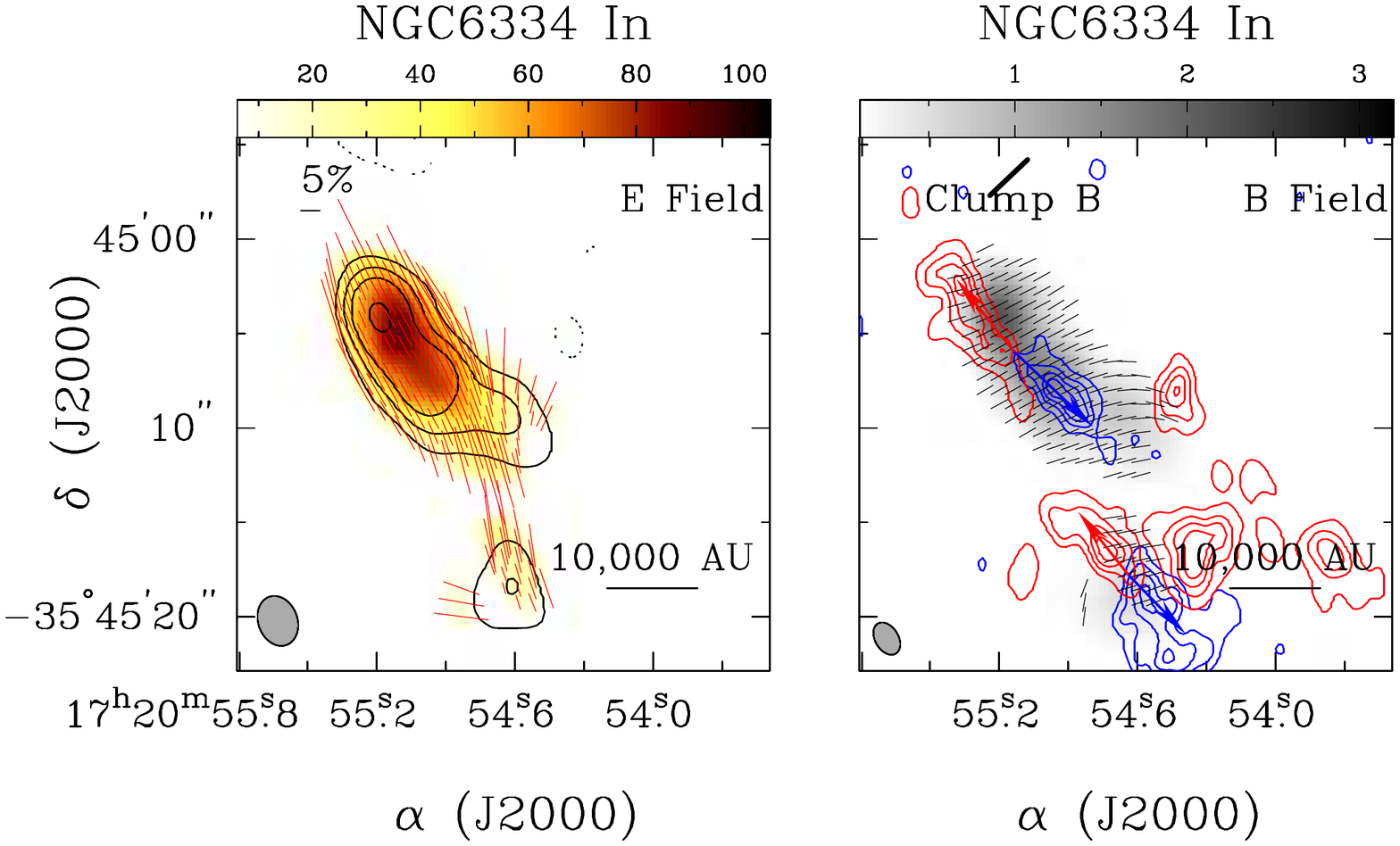}}
\caption{NGC6334In: The contour levels for the Stokes I image are $0.1 \times (\pm3, \pm6, 10, 20, 30, 40, 50, 60, 70, 80, 90, 100, 110, 120)$ Jy~beam$^{-1}$, with a synthesized beam size of $2''.71 \times 2''.07$ at a positional angle of $20^\circ$. }
\end{figure}

\pagebreak
\begin{figure}[h]
\figurenum{1f}
\resizebox{!}{12cm}{\includegraphics{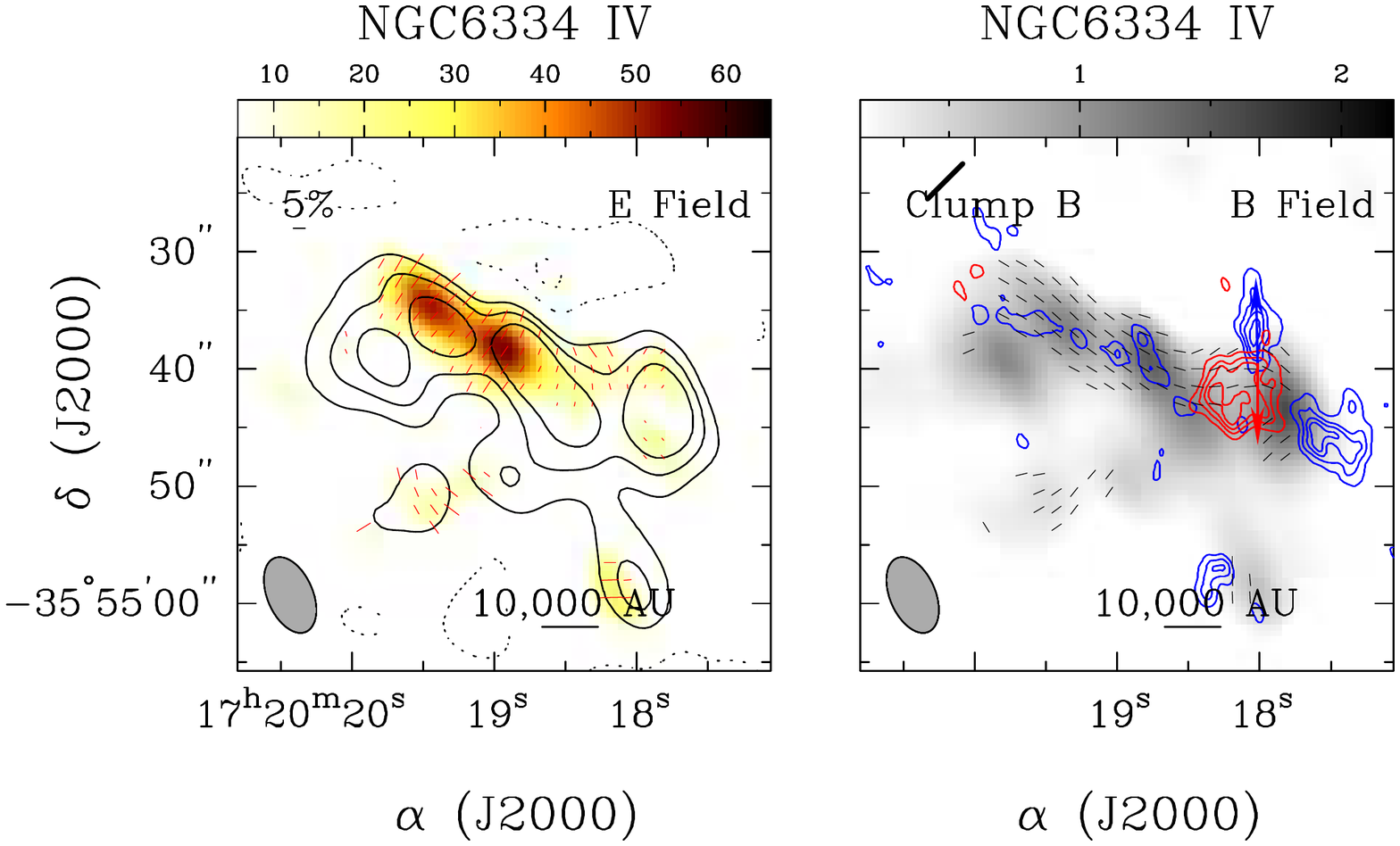}}
\caption{NGC6334 IV: The contour levels for the Stokes I image are $0.08 \times (\pm3, \pm6, 10, 20, 30, 40, 50, 60, 70, 80, 90, 100, 110, 120)$ Jy~beam$^{-1}$, with a synthesized beam size of $6''.84 \times 3''.08$ at a positional angle of $23^\circ$.  }
\end{figure}

\pagebreak
\begin{figure}[h]
\figurenum{1g}
\resizebox{!}{12cm}{\includegraphics{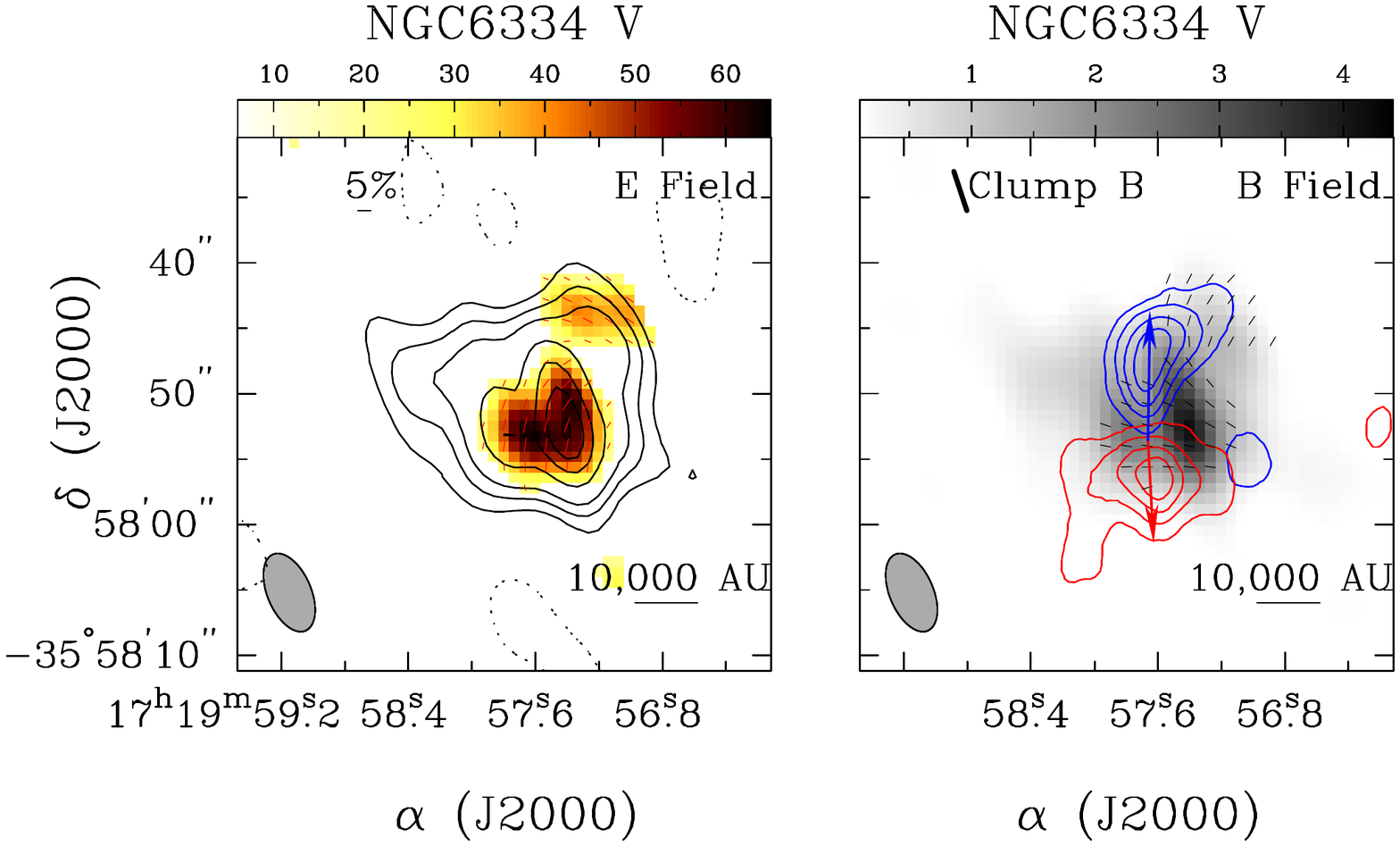}}
\caption{NGC6334 V: The contour levels for the Stokes I image are $0.08 \times (\pm3, \pm6, 10, 20, 30, 40, 50, 60, 70, 80, 90, 100, 110, 120)$ Jy~beam$^{-1}$, with a synthesized beam size of $6''.84 \times 3''.08$ at a positional angle of $23^\circ$. }
\end{figure}

\pagebreak
\begin{figure}[h]
\figurenum{1h}
\resizebox{!}{12cm}{\includegraphics{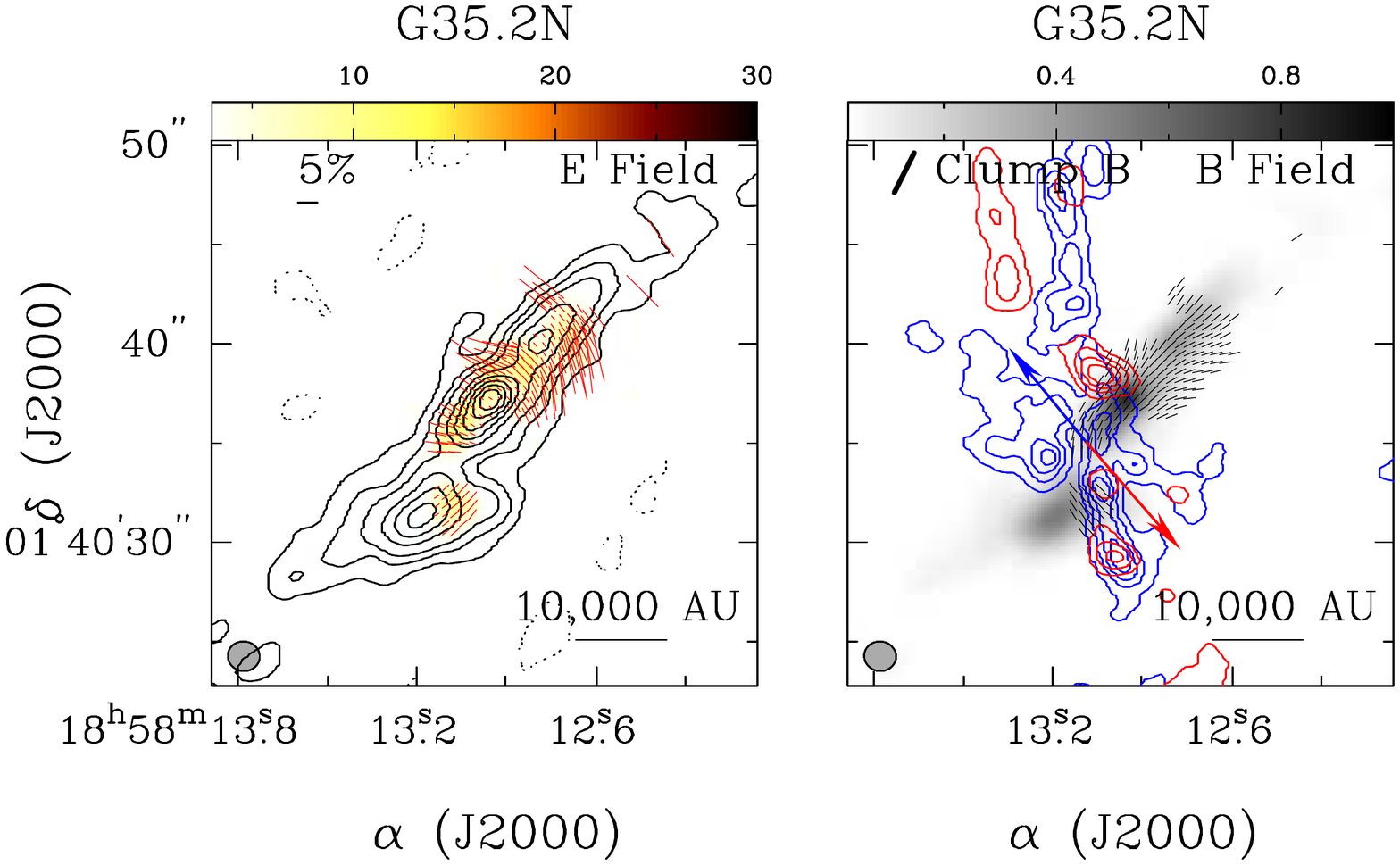}}
\caption{G35.2N: The contour levels for the Stokes I image are $0.021 \times (\pm3, \pm6, 10, 20, 30, 40, 50, 60, 70, 80, 90, 100, 110, 120)$ Jy~beam$^{-1}$, with a synthesized beam size of $1''.61 \times 1''.49$ at a positional angle of $82^\circ$. }
\end{figure}

\pagebreak
\begin{figure}[h]
\figurenum{1c}
\resizebox{!}{12cm}{\includegraphics{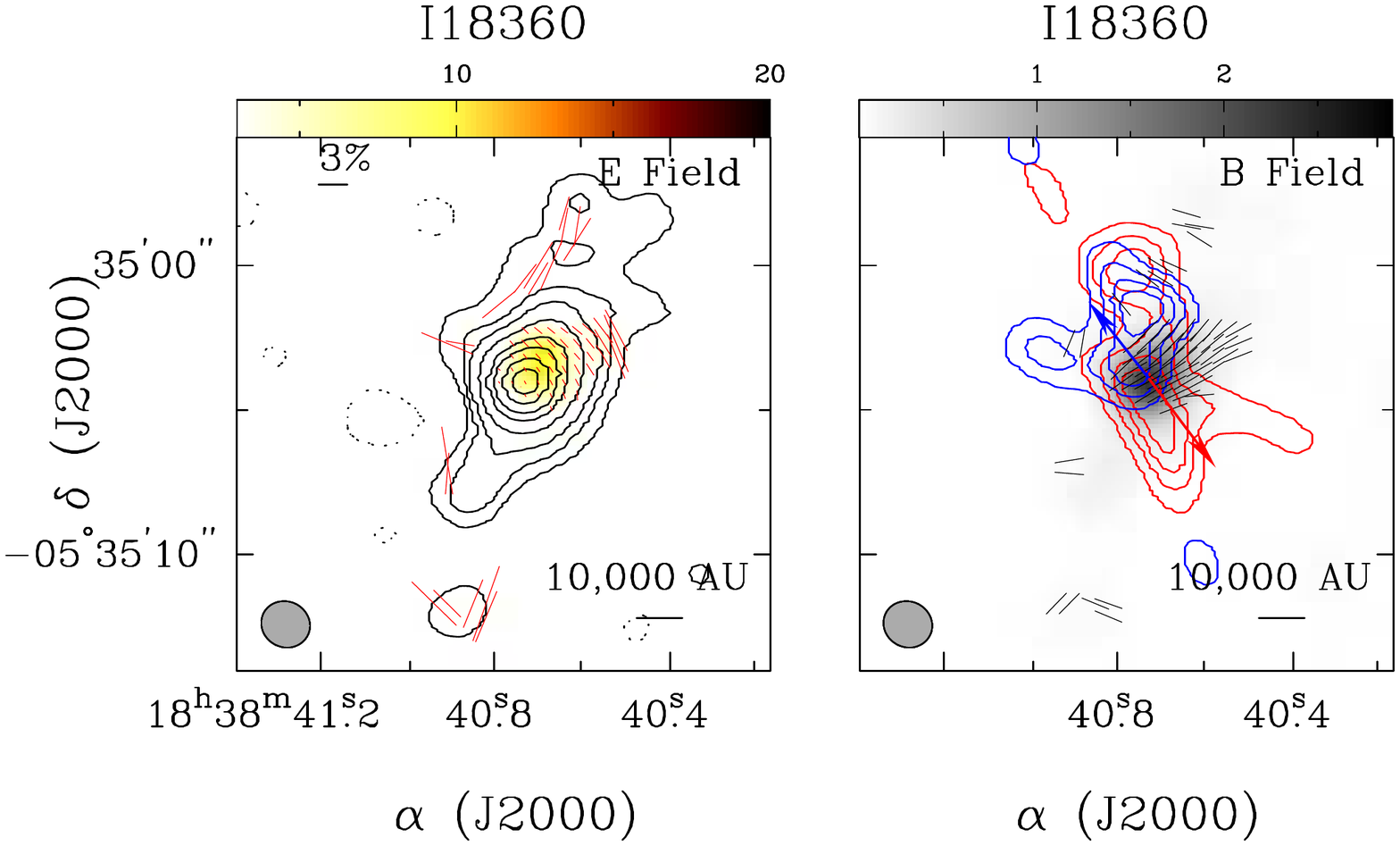}}
\caption{I18360: The contour levels for the Stokes I image are $0.017 \times (\pm3, \pm6, 10, 20, 40, 60, 90, 120, 150)$ Jy~beam$^{-1}$, with a synthesized beam size of $1''.71 \times 1''.58$  at a positional angle of $63^\circ$. }
\end{figure}

\pagebreak
\begin{figure}[h]
\figurenum{1i}
\resizebox{!}{12cm}{\includegraphics{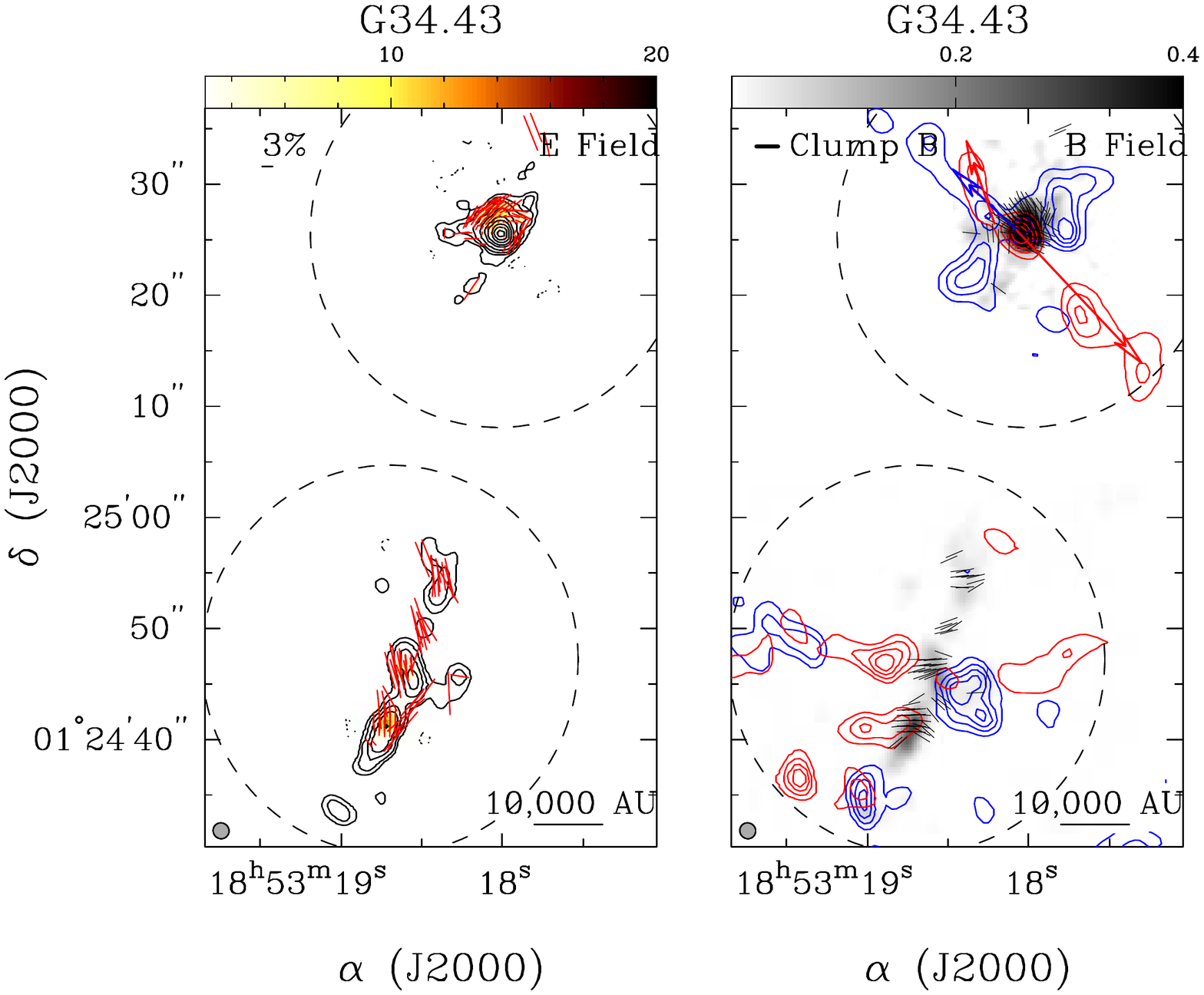}}
\caption{G34.4: The contour levels for the Stokes I image are $0.015 \times (\pm3, \pm6, 10, 20, 40, 60, 90, 120, 150)$ Jy~beam$^{-1}$, with a synthesized beam size of $1''.49 \times 1''.42$  at a positional angle of $69^\circ$. }
\end{figure}

\pagebreak
\begin{figure}[h]
\figurenum{1k}
\resizebox{!}{12cm}{\includegraphics{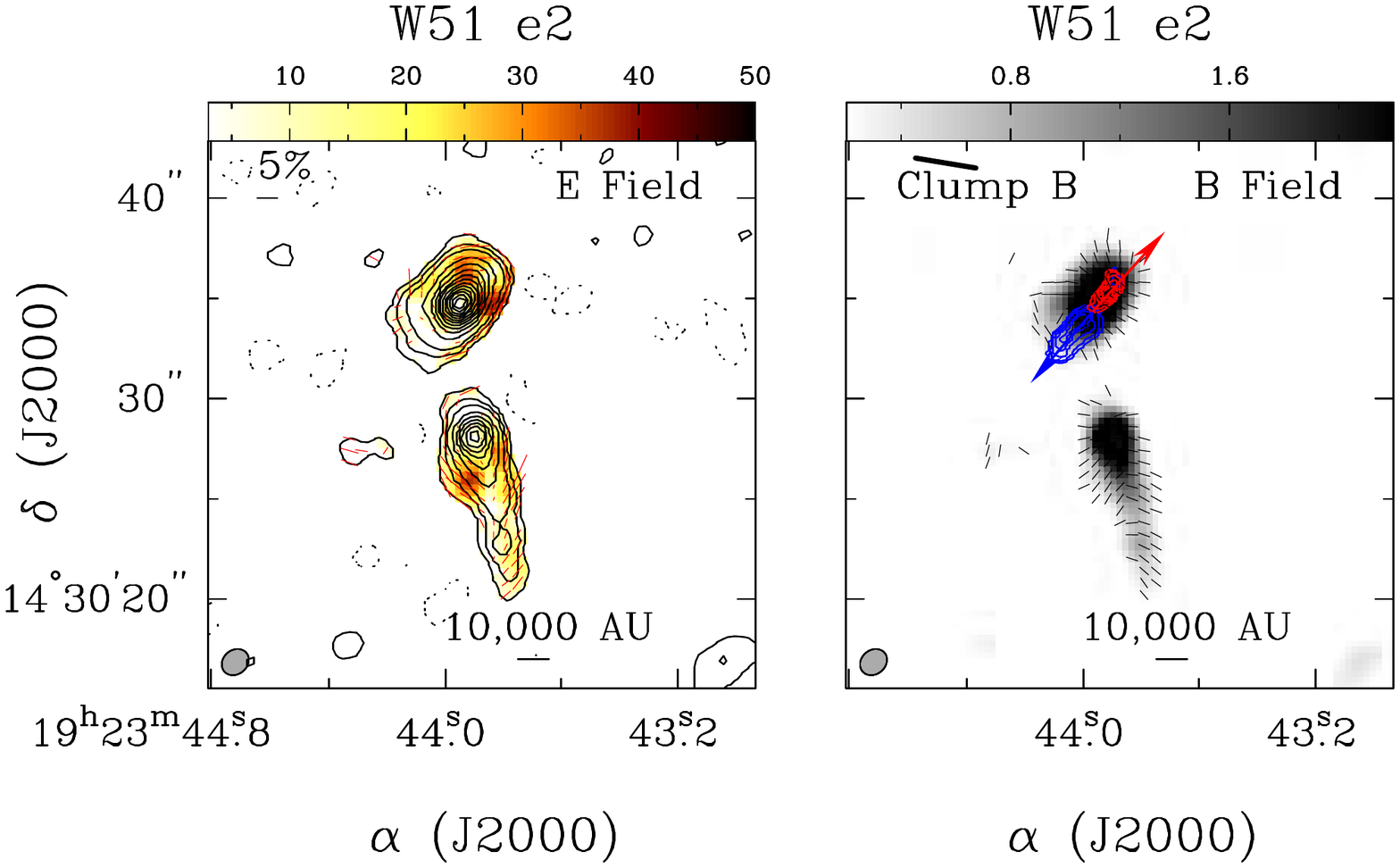}}
\caption{W51 e2: : The contour levels for the Stokes I image are $0.07 \times (\pm3, \pm6, 10, 20, 30, 40, 50, 60, 70, 80, 90, 100, 110, 120)$ Jy~beam$^{-1}$, with a synthesized beam size of $1''.44 \times 1''.20$  at a positional angle of $-50^\circ$.}
\end{figure}

\pagebreak
\begin{figure}[h]
\figurenum{1l}
\resizebox{!}{12cm}{\includegraphics{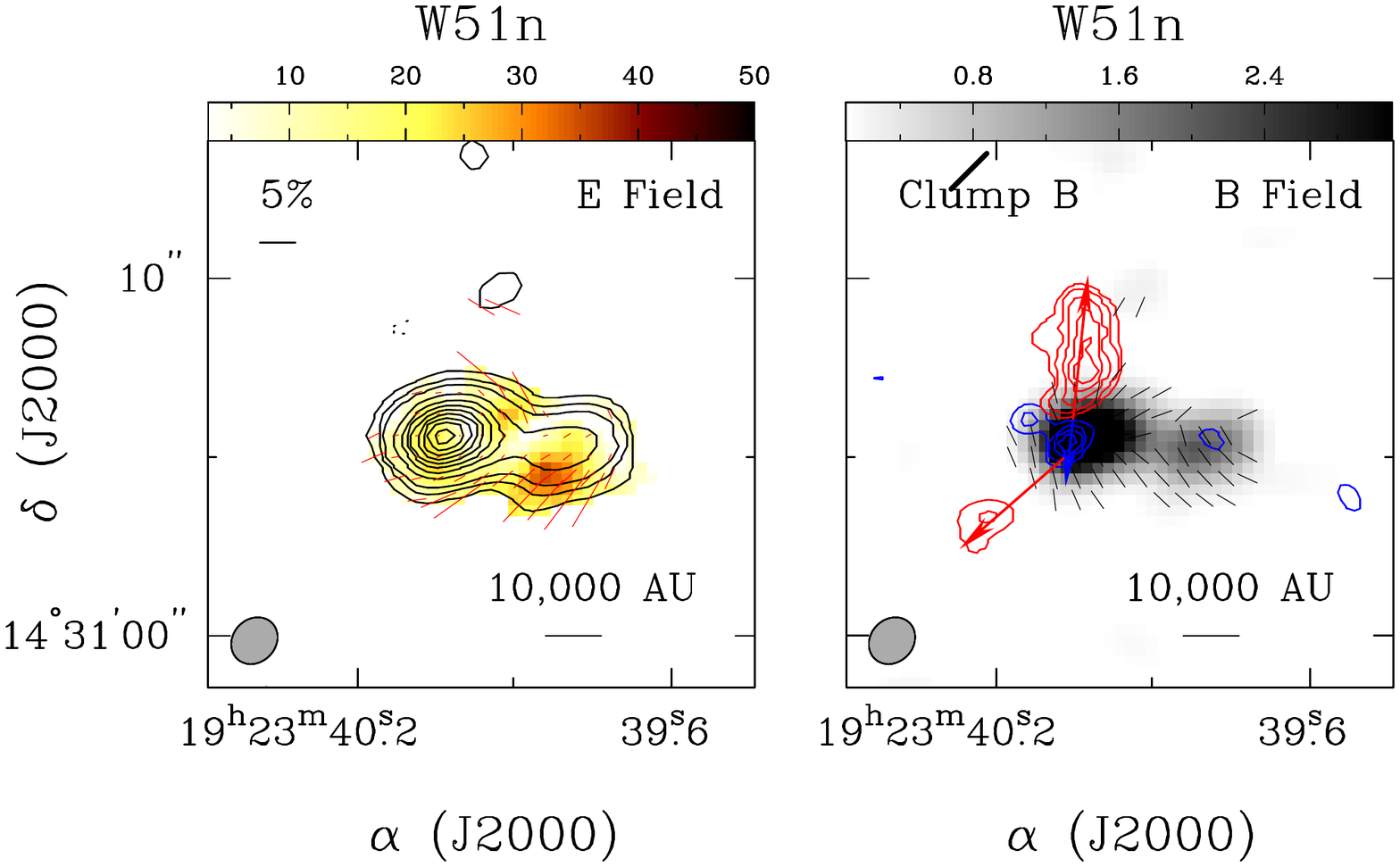}}
\caption{W51 North: The contour levels for the Stokes I image are $0.017 \times (\pm3, \pm6, 10, 20, 30, 40, 50, 60, 70, 80, 90, 100, 110, 120)$ Jy~beam$^{-1}$, with a synthesized beam size of $1''.40 \times 1''.20$  at a positional angle of $-43^\circ$.}
\end{figure}

\pagebreak
\begin{figure}[h]
\figurenum{1m}
\resizebox{!}{12cm}{\includegraphics{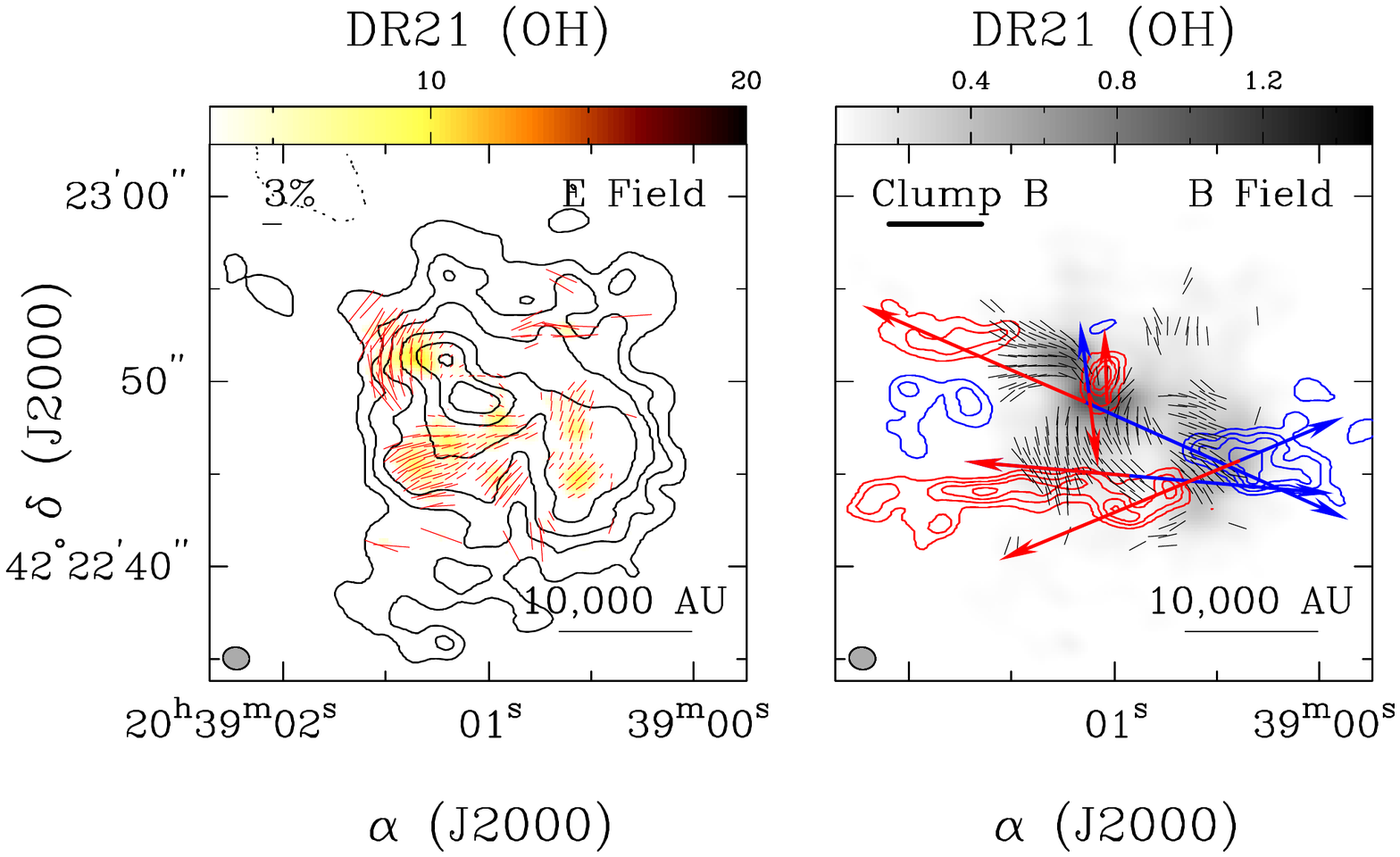}}
\caption{DR21(OH): The contour levels for the Stokes I image are $0.011 \times (\pm3, \pm6, 10, 20, 30, 40, 50, 60, 70, 80, 90, 100, 110, 120)$ Jy~beam$^{-1}$, with a synthesized beam size of $1''.43 \times 1''.21$  at a positional angle of $82^\circ$. }
\end{figure}

\pagebreak
\begin{figure}[h]
\figurenum{2}
\resizebox{!}{12cm}{\includegraphics{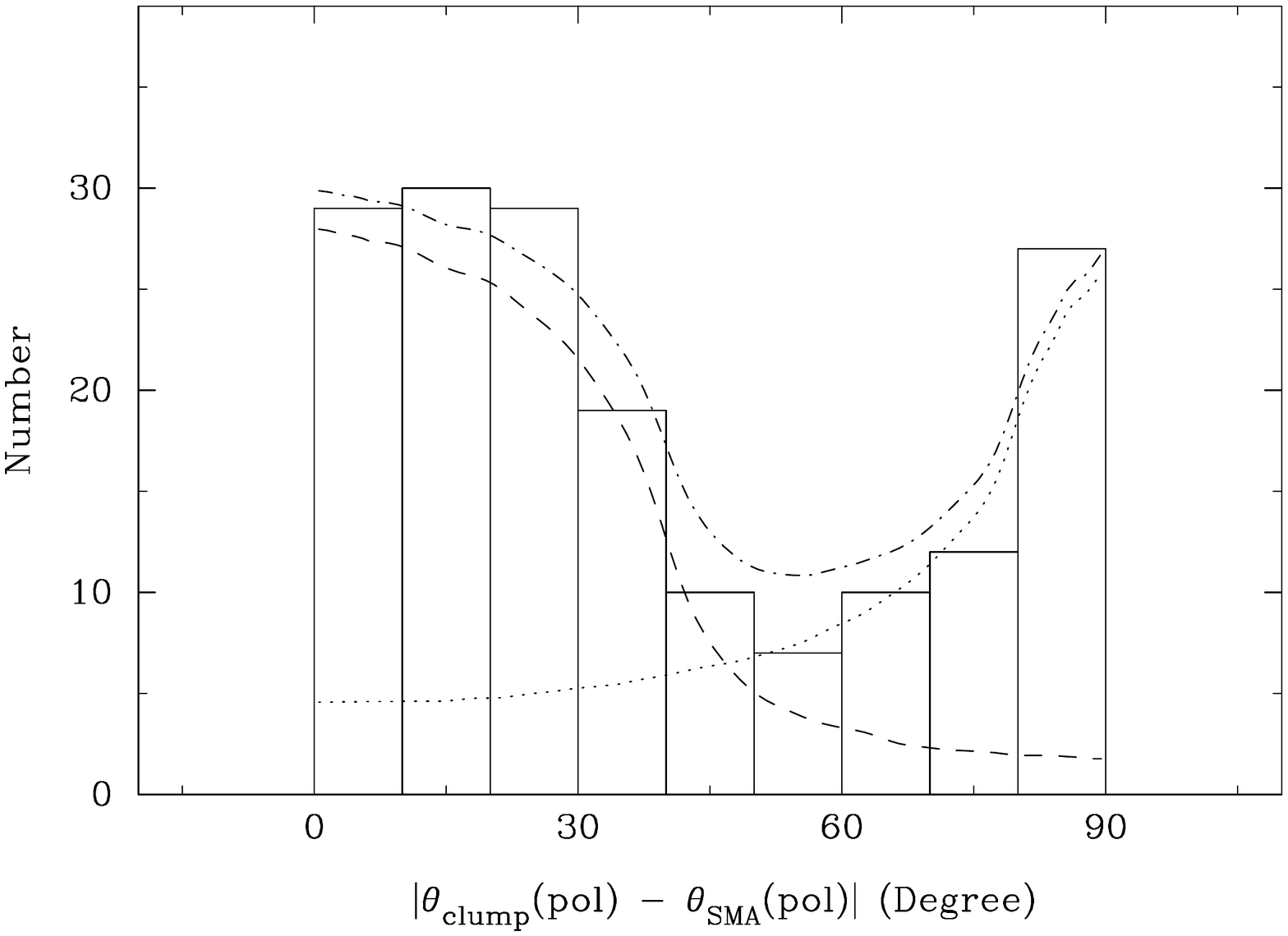}}
\caption{Difference in dust polarization angles measured at the clump scale and the core scale. 
The dashed and dotted 
lines represent the project angles of two randomly distributed vectors with an intersecting angle from 0$^\circ$ to 40$^\circ$ and from 80$^\circ$ to 90$^\circ$, respectively. The dash-dotted line is the combination of the two distributions.}
\end{figure}

\pagebreak
\begin{figure}[h]
\figurenum{3}
\resizebox{!}{12cm}{\includegraphics{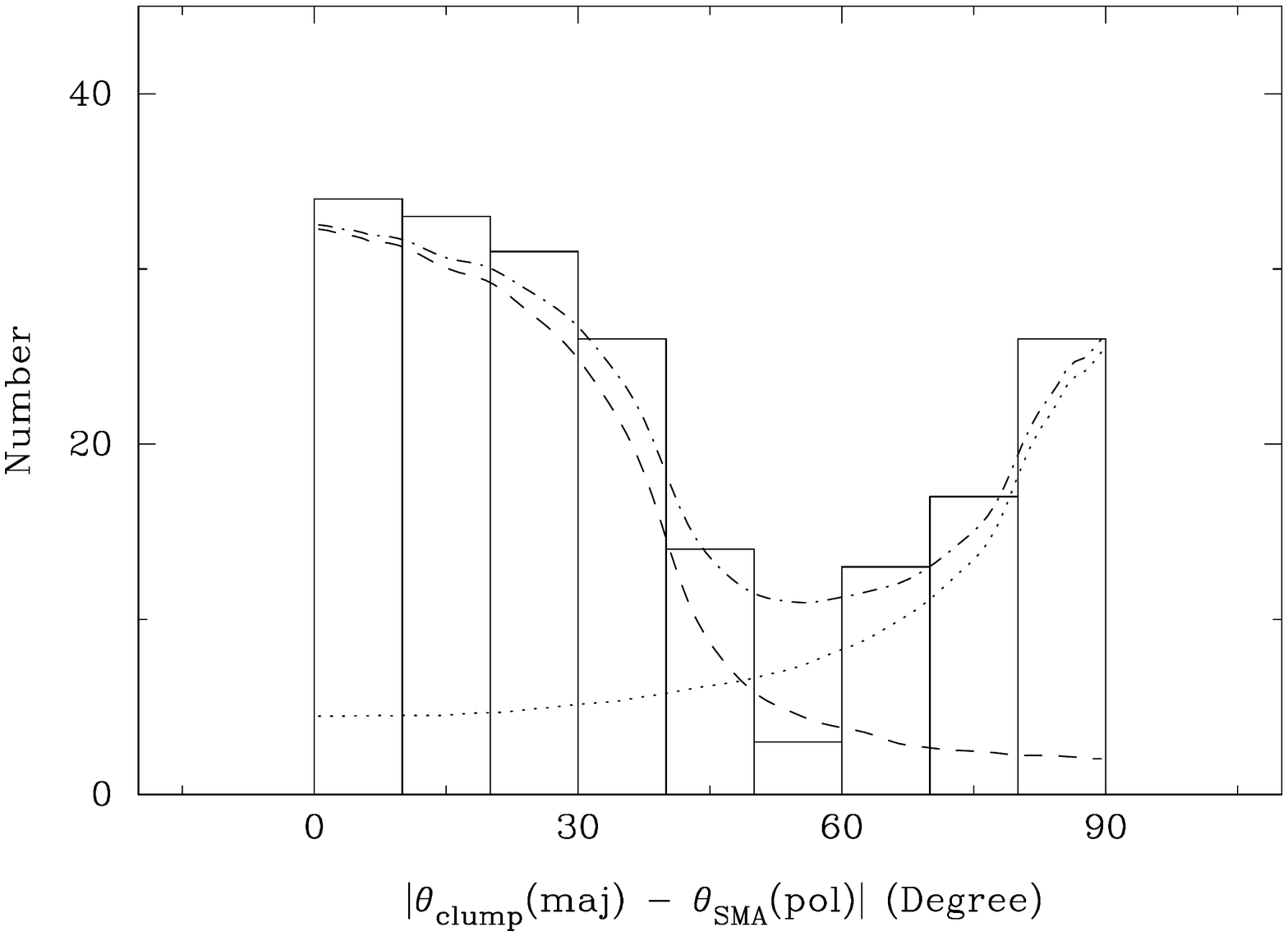}}
\caption{Difference between the major axis of the clump and dust polarization at the core scale. When the angle difference is $0^\circ$, the magnetic field is perpendicular to the long axis of the clump. The dashed and dotted 
lines represent the project angles of two randomly distributed vectors with an intersecting angle from 0$^\circ$ to 40$^\circ$ and from 80$^\circ$ to 90$^\circ$, respectively. The dash-dotted line is the combination of the two distributions.}
\end{figure}

\pagebreak
\begin{figure}[h]
\figurenum{4}
\resizebox{!}{12cm}{\includegraphics{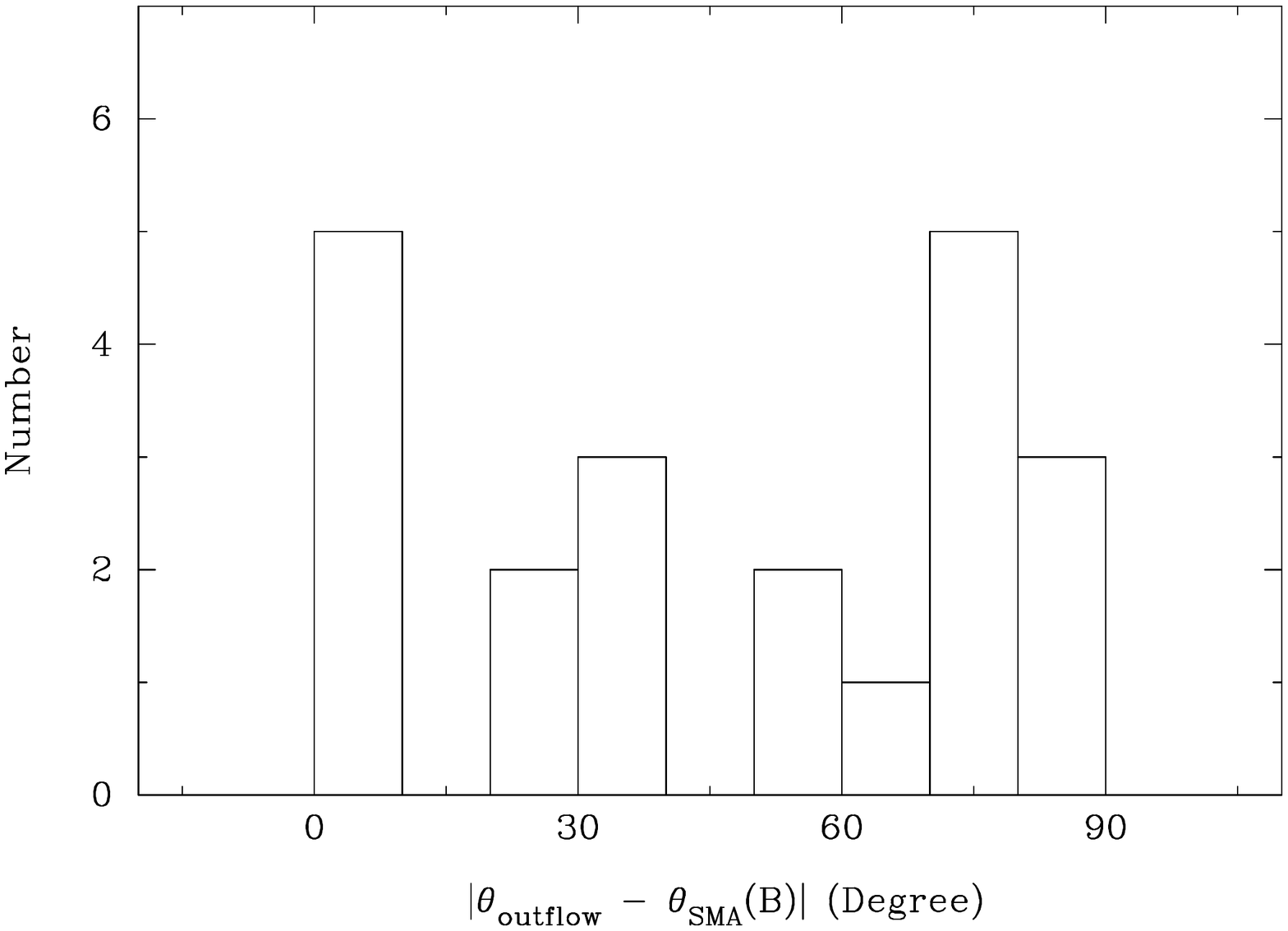}}
\caption{Difference between the position angle of the outflow axis and magnetic fields in the core where outflows originate. When the angle difference is $0^\circ$,  the major axis of the outflow is aligned with the core magnetic field.}
\end{figure}

\pagebreak
\begin{figure}[h]
\figurenum{5}
\resizebox{!}{12cm}{\includegraphics{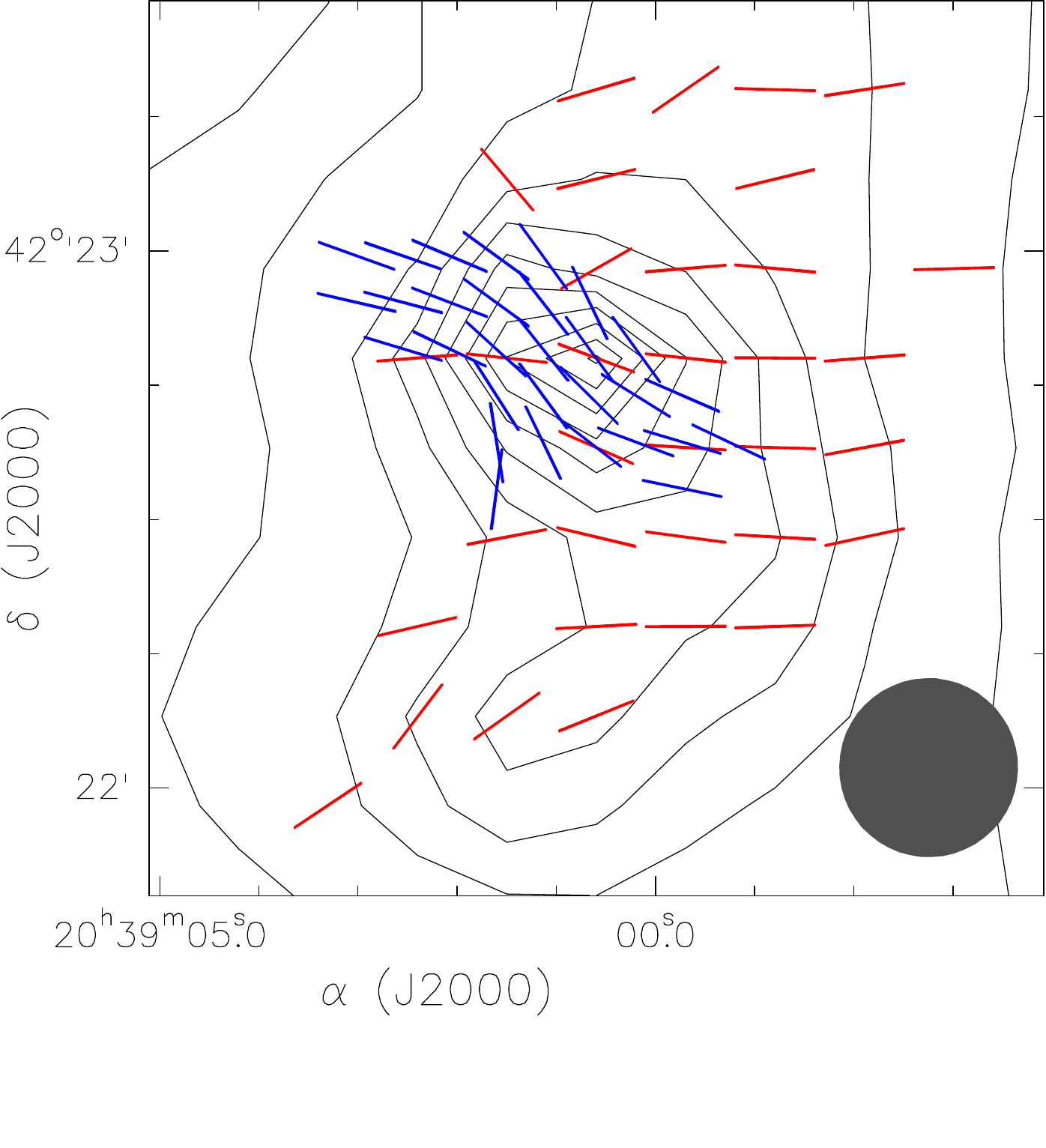}}
\caption{Contour map of the dust emission at 870 $\mu$m towards the DR21 region,
overlaid with the magnetic field segments (red bars) obtained with the JCMT SCUBA polarimeter \citep{matthews2009}.
The blue line segments represent the magnetic field direction obtained with the SMA, but convolved to
20$''$, the angular resolution of the SCUBA data. The shaded circle at the lower-right corner of the panel marks the 20$''$ beam.}
\end{figure}

\pagebreak
\begin{figure}[h]
\figurenum{6}
\hspace{8.3cm}
\resizebox{!}{8cm}{\includegraphics{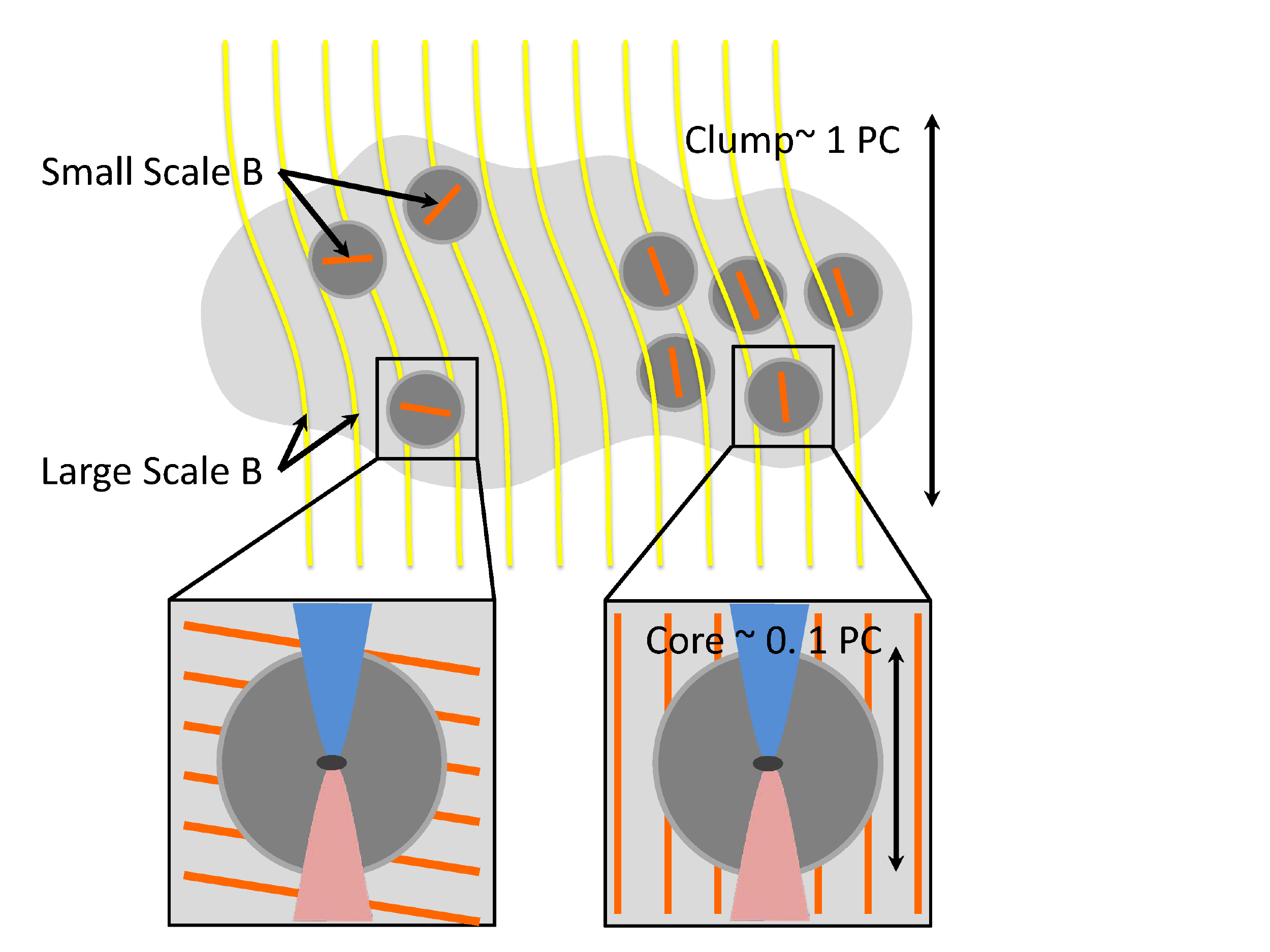}}
\caption{A schematic picture demonstrating the dynamical role of magnetic fields in the formation of dense cores within a parsec-scale molecular clump, and at scales of accretion disks. Magnetic fields in core scales are either parallel or perpendicular to the clump-scale field, indicating that they play an important dynamical role in dense core formation. On the other hand, outflow axes are not aligned with the field in dense cores, indicating that angular momenta in disks are not aligned with the core field.}
\end{figure}


\newpage

\begin{thebibliography}{87}
\expandafter\ifx\csname natexlab\endcsname\relax\def\natexlab#1{#1}\fi

\bibitem[{{Allen} {et~al.}(2003){Allen}, {Shu}, \& {Li}}]{allen2003a}
{Allen}, A., {Shu}, F.~H., \& {Li}, Z.-Y. 2003, \apj, 599, 351


\bibitem[{{Beuther} {et~al.}(2002{\natexlab{a}}){Beuther}, {Schilke}, {Gueth},
  {McCaughrean}, {Andersen}, {Sridharan}, \& {Menten}}]{beuther2002d}
{Beuther}, H., {Schilke}, P., {Gueth}, F., {et~al.} 2002{\natexlab{a}}, \aap,
  387, 931

\bibitem[{{Beuther} {et~al.}(2002{\natexlab{b}}){Beuther}, {Schilke}, {Menten},
  {Motte}, {Sridharan}, \& {Wyrowski}}]{beuther2002a}
{Beuther}, H., {Schilke}, P., {Menten}, K.~M., {et~al.} 2002{\natexlab{b}},
  \apj, 566, 945

\bibitem[{{Beuther} {et~al.}(2002{\natexlab{c}}){Beuther}, {Schilke},
  {Sridharan}, {Menten}, {Walmsley}, \& {Wyrowski}}]{beuther2002b}
{Beuther}, H., {Schilke}, P., {Sridharan}, T.~K., {et~al.} 2002{\natexlab{c}},
  \aap, 383, 892

\bibitem[{{Busquet} {et~al.}(2013){Busquet}, {Zhang}, {Palau}, {Liu},
  {S{\'a}nchez-Monge}, {Estalella}, {Ho}, {de Gregorio-Monsalvo}, {Pillai},
  {Wyrowski}, {Girart}, {Santos}, \& {Franco}}]{busquet2013}
{Busquet}, G., {Zhang}, Q., {Palau}, A., {et~al.} 2013, \apjl, 764, L26

\bibitem[{{Cesaroni} {et~al.}(1999){Cesaroni}, {Felli}, {Jenness}, {Neri},
  {Olmi}, {Robberto}, {Testi}, \& {Walmsley}}]{cesaroni1999a}
{Cesaroni}, R., {Felli}, M., {Jenness}, T., {et~al.} 1999, \aap, 345, 949

\bibitem[{{Cesaroni} {et~al.}(2007){Cesaroni}, {Galli}, {Lodato}, {Walmsley},
  \& {Zhang}}]{cesaroni2007}
{Cesaroni}, R., {Galli}, D., {Lodato}, G., {Walmsley}, C.~M., \& {Zhang}, Q.
  2007, in Protostars and Planets V, ed. B.~{Reipurth}, D.~{Jewitt}, \&
  K.~{Keil}, 197--212

\bibitem[{{Chapman} {et~al.}(2013){Chapman}, {Davidson}, {Goldsmith}, {Houde},
  {Kwon}, {Li}, {Looney}, {Matthews}, {Matthews}, {Novak}, {Peng},
  {Vaillancourt}, \& {Volgenau}}]{chapman2013}
{Chapman}, N.~L., {Davidson}, J.~A., {Goldsmith}, P.~F., {et~al.} 2013, \apj,
  770, 151

\bibitem[{{Cortes} {et~al.}(2008){Cortes}, {Crutcher}, {Shepherd}, \&
  {Bronfman}}]{cortes2008}
{Cortes}, P.~C., {Crutcher}, R.~M., {Shepherd}, D.~S., \& {Bronfman}, L. 2008,
  \apj, 676, 464

\bibitem[{{Crutcher}(2012)}]{crutcher2012}
{Crutcher}, R.~M. 2012, \araa, 50, 29

\bibitem[{{Csengeri} {et~al.}(2011){Csengeri}, {Bontemps}, {Schneider},
  {Motte}, \& {Dib}}]{csengeri2011}
{Csengeri}, T., {Bontemps}, S., {Schneider}, N., {Motte}, F., \& {Dib}, S.
  2011, \aap, 527, A135+

\bibitem[{{Dotson} {et~al.}(2010){Dotson}, {Vaillancourt}, {Kirby}, {Dowell},
  {Hildebrand}, \& {Davidson}}]{dotson2010}
{Dotson}, J.~L., {Vaillancourt}, J.~E., {Kirby}, L., {et~al.} 2010, \apjs, 186,
  406

\bibitem[{{Duarte-Cabral} {et~al.}(2013){Duarte-Cabral}, {Bontemps}, {Motte},
  {Hennemann}, {Schneider}, \& {Andr{\'e}}}]{duarte-cabral2013}
{Duarte-Cabral}, A., {Bontemps}, S., {Motte}, F., {et~al.} 2013, \aap, 558,
  A125

\bibitem[{{Falgarone} {et~al.}(2008){Falgarone}, {Troland}, {Crutcher}, \&
  {Paubert}}]{falgarone2008}
{Falgarone}, E., {Troland}, T.~H., {Crutcher}, R.~M., \& {Paubert}, G. 2008,
  \aap, 487, 247

\bibitem[{{Franco} {et~al.}(2010){Franco}, {Alves}, \& {Girart}}]{franco2010}
{Franco}, G.~A.~P., {Alves}, F.~O., \& {Girart}, J.~M. 2010, \apj, 723, 146

\bibitem[{{Frau} {et~al.}(2011){Frau}, {Galli}, \& {Girart}}]{frau2011}
{Frau}, P., {Galli}, D., \& {Girart}, J.~M. 2011, \aap, 535, A44

\bibitem[{{Frau} {et~al.}(2014){Frau}, {Girart}, {Zhang}, \& {Rao}}]{frau2014}
{Frau}, P., {Girart}, J.~M., {Zhang}, Q., \& {Rao}, R. 2014, submitted to \apj

\bibitem[{{Galli} \& {Shu}(1993)}]{galli1993}
{Galli}, D. \& {Shu}, F.~H. 1993, \apj, 417, 220

\bibitem[{{Girart} {et~al.}(2009){Girart}, {Beltr{\'a}n}, {Zhang}, {Rao}, \&
  {Estalella}}]{girart2009}
{Girart}, J.~M., {Beltr{\'a}n}, M.~T., {Zhang}, Q., {Rao}, R., \& {Estalella},
  R. 2009, Science, 324, 1408

\bibitem[{{Girart} {et~al.}(2013){Girart}, {Frau}, {Zhang}, {Koch}, {Qiu},
  {Tang}, {Lai}, \& {Ho}}]{girart2013}
{Girart}, J.~M., {Frau}, P., {Zhang}, Q., {et~al.} 2013, \apj, 772, 69

\bibitem[{{Girart} {et~al.}(2006){Girart}, {Rao}, \& {Marrone}}]{girart2006}
{Girart}, J.~M., {Rao}, R., \& {Marrone}, D.~P. 2006, Science, 313, 812

\bibitem[{{Gon{\c c}alves} {et~al.}(2008){Gon{\c c}alves}, {Galli}, \&
  {Girart}}]{goncalves2008}
{Gon{\c c}alves}, J., {Galli}, D., \& {Girart}, J.~M. 2008, \aap, 490, L39

\bibitem[{{Goodman} {et~al.}(1998){Goodman}, {Barranco}, {Wilner}, \&
  {Heyer}}]{goodman1998}
{Goodman}, A.~A., {Barranco}, J.~A., {Wilner}, D.~J., \& {Heyer}, M.~H. 1998,
  \apj, 504, 223

\bibitem[{{Heyer} {et~al.}(2008){Heyer}, {Gong}, {Ostriker}, \&
  {Brunt}}]{heyer2008}
{Heyer}, M., {Gong}, H., {Ostriker}, E., \& {Brunt}, C. 2008, \apj, 680, 420

\bibitem[{{Ho} {et~al.}(2004){Ho}, {Moran}, \& {Lo}}]{ho2004}
{Ho}, P.~T.~P., {Moran}, J.~M., \& {Lo}, K.~Y. 2004, \apjl, 616, L1

\bibitem[{{Houde} {et~al.}(2004){Houde}, {Dowell}, {Hildebrand}, {Dotson},
  {Vaillancourt}, {Phillips}, {Peng}, \& {Bastien}}]{houde2004}
{Houde}, M., {Dowell}, C.~D., {Hildebrand}, R.~H., {et~al.} 2004, \apj, 604,
  717

\bibitem[{{Hull} {et~al.}(2013){Hull}, {Plambeck}, {Bolatto}, {Bower},
  {Carpenter}, {Crutcher}, {Fiege}, {Franzmann}, {Hakobian}, {Heiles}, {Houde},
  {Hughes}, {Jameson}, {Kwon}, {Lamb}, {Looney}, {Matthews}, {Mundy}, {Pillai},
  {Pound}, {Stephens}, {Tobin}, {Vaillancourt}, {Volgenau}, \&
  {Wright}}]{hull2013}
{Hull}, C.~L.~H., {Plambeck}, R.~L., {Bolatto}, A.~D., {et~al.} 2013, \apj,
  768, 159

\bibitem[{{Jammalamadaka \& Sengupta}(2001){Jammalamadaka}, \& {Sengupta}}]{Jammalamadaka2001}
Jammalamadaka S. R., \& Sengupta A 2001, Topics in Circular Statistics. World Scientific

\bibitem[{{Jijina} {et~al.}(1999){Jijina}, {Myers}, \& {Adams}}]{jijina1999}
{Jijina}, J., {Myers}, P.~C., \& {Adams}, F.~C. 1999, \apjs, 125, 161

\bibitem[{{Keto} \& {Zhang}(2010)}]{keto2010}
{Keto}, E. \& {Zhang}, Q. 2010, \mnras, 406, 102

\bibitem[{{Lada} \& {Lada}(2003)}]{lada2003}
{Lada}, C.~J. \& {Lada}, E.~A. 2003, \araa, 41, 57

\bibitem[{{Lai} {et~al.}(2001){Lai}, {Crutcher}, {Girart}, \& {Rao}}]{lai2001}
{Lai}, S.-P., {Crutcher}, R.~M., {Girart}, J.~M., \& {Rao}, R. 2001, \apj, 561,
  864

\bibitem[{{Lazarian}(2007)}]{lazarian2007}
{Lazarian}, A. 2007, \jqsrt, 106, 225

\bibitem[{{Li} {et~al.}(2013){Li}, {Fang}, {Henning}, \&
  {Kainulainen}}]{li2013}
{Li}, H.-b., {Fang}, M., {Henning}, T., \& {Kainulainen}, J. 2013, \mnras, 436,
  3707

\bibitem[{{Li} {et~al.}(2014){Li}, Goodman, Sridharan, Houde, Li, Novak, \&
  Tang}]{li2014}
{Li}, H.-b., Goodman, A., Sridharan, T., {et~al.} 2014, in accepted to
  Protostars and Planets VI, ed. H.~Beuther, R.~Klessen, C.~Dullemond, \&
  H.~Th.

\bibitem[{{Liu} {et~al.}(2013){Liu}, {Qiu}, {Zhang}, {Girart}, \&
  {Ho}}]{liu2013}
{Liu}, H.~B., {Qiu}, K., {Zhang}, Q., {Girart}, J.~M., \& {Ho}, P.~T.~P. 2013,
  \apj, 771, 71

\bibitem[{Lu {et~al.}(2014)Lu, Zhang, Liu, Wang, \& Gu}]{lu2014}
Lu, X., Zhang, Q., Liu, H.~B., Wang, J., \& Gu, Q. 2014, \apj accepted

\bibitem[{{Marrone}(2006)}]{marrone2006}
{Marrone}, D.~P. 2006, PhD thesis, Harvard University

\bibitem[{{Marrone} \& {Rao}(2008)}]{marrone2008}
{Marrone}, D.~P. \& {Rao}, R. 2008, in Society of Photo-Optical Instrumentation
  Engineers (SPIE) Conference Series, Vol. 7020, Society of Photo-Optical
  Instrumentation Engineers (SPIE) Conference Series

\bibitem[{{Marti} {et~al.}(1998){Marti}, {Rodriguez}, \&
  {Reipurth}}]{marti1998}
{Marti}, J., {Rodriguez}, L.~F., \& {Reipurth}, B. 1998, \apj, 502, 337

\bibitem[{{Matthews} {et~al.}(2009){Matthews}, {McPhee}, {Fissel}, \&
  {Curran}}]{matthews2009}
{Matthews}, B.~C., {McPhee}, C.~A., {Fissel}, L.~M., \& {Curran}, R.~L. 2009,
  \apjs, 182, 143

\bibitem[{{Mellon} \& {Li}(2009)}]{mellon2009}
{Mellon}, R.~R. \& {Li}, Z.-Y. 2009, \apj, 698, 922

\bibitem[{{Motte} {et~al.}(2007){Motte}, {Bontemps}, {Schilke}, {Schneider},
  {Menten}, \& {Brogui{\`e}re}}]{motte2007}
{Motte}, F., {Bontemps}, S., {Schilke}, P., {et~al.} 2007, \aap, 476, 1243

\bibitem[{{Mouschovias}(1976)}]{mouschovias1976}
{Mouschovias}, T.~C. 1976, \apj, 207, 141

\bibitem[{{Mu{\~n}oz} {et~al.}(2007){Mu{\~n}oz}, {Mardones}, {Garay},
  {Rebolledo}, {Brooks}, \& {Bontemps}}]{munoz2007}
{Mu{\~n}oz}, D.~J., {Mardones}, D., {Garay}, G., {et~al.} 2007, \apj, 668, 906

\bibitem[{{Myers} \& {Benson}(1983)}]{myers1983}
{Myers}, P.~C. \& {Benson}, P.~J. 1983, \apj, 266, 309

\bibitem[{{Nagai} {et~al.}(1998){Nagai}, {Inutsuka}, \& {Miyama}}]{nagai1998}
{Nagai}, T., {Inutsuka}, S.-I., \& {Miyama}, S.~M. 1998, \apj, 506, 306

\bibitem[{{Naghizadeh-Khouei} \& {Clarke}(1993)}]{naghizadeh1993}
{Naghizadeh-Khouei}, J. \& {Clarke}, D. 1993, \aap, 274, 968

\bibitem[{{Nakamura} \& {Li}(2008)}]{nakamura2008}
{Nakamura}, F. \& {Li}, Z.-Y. 2008, \apj, 687, 354

\bibitem[{{Ostriker} {et~al.}(2001){Ostriker}, {Stone}, \&
  {Gammie}}]{ostriker2001}
{Ostriker}, E.~C., {Stone}, J.~M., \& {Gammie}, C.~F. 2001, \apj, 546, 980

\bibitem[{{Padovani} {et~al.}(2013){Padovani}, {Hennebelle}, \&
  {Galli}}]{padovani2013}
{Padovani}, M., {Hennebelle}, P., \& {Galli}, D. 2013, \aap, 560, A114

\bibitem[{{Pillai} {et~al.}(2006){Pillai}, {Wyrowski}, {Carey}, \&
  {Menten}}]{pillai2006b}
{Pillai}, T., {Wyrowski}, F., {Carey}, S.~J., \& {Menten}, K.~M. 2006, \aap,
  450, 569

\bibitem[{{Qiu} \& {Zhang}(2009)}]{qiu2009b}
{Qiu}, K. \& {Zhang}, Q. 2009, \apjl, 702, L66

\bibitem[{{Qiu} {et~al.}(2007){Qiu}, {Zhang}, {Beuther}, \& {Yang}}]{qiu2007}
{Qiu}, K., {Zhang}, Q., {Beuther}, H., \& {Yang}, J. 2007, \apj, 654, 361

\bibitem[{{Qiu} {et~al.}(2008){Qiu}, {Zhang}, {Megeath}, {Gutermuth},
  {Beuther}, {Shepherd}, {Sridharan}, {Testi}, \& {De Pree}}]{qiu2008}
{Qiu}, K., {Zhang}, Q., {Megeath}, S.~T., {et~al.} 2008, ArXiv e-prints, 806

\bibitem[{{Qiu} {et~al.}(2011){Qiu}, {Zhang}, \& {Menten}}]{qiu2011}
{Qiu}, K., {Zhang}, Q., \& {Menten}, K.~M. 2011, \apj, 728, 6

\bibitem[{{Qiu} {et~al.}(2013){Qiu}, {Zhang}, {Menten}, {Liu}, \&
  {Tang}}]{qiu2013}
{Qiu}, K., {Zhang}, Q., {Menten}, K.~M., {Liu}, H.~B., \& {Tang}, Y.-W. 2013,
  ArXiv e-prints

\bibitem[{{Qiu} {et~al.}(2009){Qiu}, {Zhang}, {Wu}, \& {Chen}}]{qiu2009a}
{Qiu}, K., {Zhang}, Q., {Wu}, J., \& {Chen}, H.-R. 2009, \apj, 696, 66

\bibitem[{{Rao} {et~al.}(1998){Rao}, {Crutcher}, {Plambeck}, \&
  {Wright}}]{rao1998}
{Rao}, R., {Crutcher}, R.~M., {Plambeck}, R.~L., \& {Wright}, M.~C.~H. 1998,
  \apjl, 502, L75

\bibitem[{{Rao} {et~al.}(2009){Rao}, {Girart}, {Marrone}, {Lai}, \&
  {Schnee}}]{rao2009}
{Rao}, R., {Girart}, J.~M., {Marrone}, D.~P., {Lai}, S.-P., \& {Schnee}, S.
  2009, \apj, 707, 921

\bibitem[{{Rathborne} {et~al.}(2006){Rathborne}, {Jackson}, \&
  {Simon}}]{rathborne2006}
{Rathborne}, J.~M., {Jackson}, J.~M., \& {Simon}, R. 2006, \apj, 641, 389

\bibitem[{{Russeil} {et~al.}(2013){Russeil}, {Schneider}, {Anderson},
  {Zavagno}, {Molinari}, {Persi}, {Bontemps}, {Motte}, {Ossenkopf},
  {Andr{\'e}}, {Arzoumanian}, {Bernard}, {Deharveng}, {Didelon}, {Di
  Francesco}, {Elia}, {Hennemann}, {Hill}, {K{\"o}nyves}, {Li}, {Martin},
  {Nguyen Luong}, {Peretto}, {Pezzuto}, {Polychroni}, {Roussel}, {Rygl},
  {Spinoglio}, {Testi}, {Tig{\'e}}, {Vavrek}, {Ward-Thompson}, \&
  {White}}]{russeil2013}
{Russeil}, D., {Schneider}, N., {Anderson}, L.~D., {et~al.} 2013, \aap, 554,
  A42

\bibitem[{{S{\'a}nchez-Monge} {et~al.}(2013){S{\'a}nchez-Monge}, {Palau},
  {Fontani}, {Busquet}, {Ju{\'a}rez}, {Estalella}, {Tan}, {Sep{\'u}lveda},
  {Ho}, {Zhang}, \& {Kurtz}}]{sanchez2013a}
{S{\'a}nchez-Monge}, {\'A}., {Palau}, A., {Fontani}, F., {et~al.} 2013, \mnras,
  432, 3288

\bibitem[{{Sandell}(2000)}]{sandell2000}
{Sandell}, G. 2000, \aap, 358, 242

\bibitem[{{Sault} {et~al.}(1995){Sault}, {Teuben}, \& {Wright}}]{sault1995}
{Sault}, R.~J., {Teuben}, P.~J., \& {Wright}, M.~C.~H. 1995, in ASP Conf. Ser.
  77: Astronomical Data Analysis Software and Systems IV, 433

\bibitem[{{Serkowski}(1974)}]{serkowski1974}
{Serkowski}, K. 1974, {Polarization techniques.}, ed. N.~P. {Carleton},
  361--414

\bibitem[{{Shang} {et~al.}(2007){Shang}, {Li}, \& {Hirano}}]{shang2007}
{Shang}, H., {Li}, Z.-Y., \& {Hirano}, N. 2007, Protostars and Planets V, 261

\bibitem[{{Shi}(2010)}]{shi2010}
{Shi}, H., {Zhao}, J.-H., \& {Han}, J.~L. 2010, \apjl, 718, L181

\bibitem[{{Shu}(1977)}]{shu1977}
{Shu}, F.~H. 1977, \apj, 214, 488

\bibitem[{{Shu} {et~al.}(1987){Shu}, {Adams}, \& {Lizano}}]{shu1987}
{Shu}, F.~H., {Adams}, F.~C., \& {Lizano}, S. 1987, \araa, 25, 23

\bibitem[{{Sollins} {et~al.}(2004){Sollins}, {Hunter}, {Battat}, {Beuther},
  {Ho}, {Lim}, {Liu}, {Ohashi}, {Sridharan}, {Su}, {Zhao}, \&
  {Zhang}}]{sollins2004b}
{Sollins}, P.~K., {Hunter}, T.~R., {Battat}, J., {et~al.} 2004, \apjl, 616, L35

\bibitem[{{Stephens} {et~al.}(2013){Stephens}, {Looney}, {Kwon}, {Hull},
  {Plambeck}, {Crutcher}, {Chapman}, {Novak}, {Davidson}, {Vaillancourt},
  {Shinnaga}, \& {Matthews}}]{stephens2013}
{Stephens}, I.~W., {Looney}, L.~W., {Kwon}, W., {et~al.} 2013, \apjl, 769, L15

\bibitem[{{Su} {et~al.}(2004){Su}, {Zhang}, \& {Lim}}]{su2004}
{Su}, Y., {Zhang}, Q., \& {Lim}, J. 2004, \apj, 604, 258

\bibitem[{{Su} {et~al.}(2007){Su}, {Liu}, {Chen}, {Zhang}, \&
  {Cesaroni}}]{su2007}
{Su}, Y.-N., {Liu}, S.-Y., {Chen}, H.-R., {Zhang}, Q., \& {Cesaroni}, R. 2007,
  \apj, 671, 571

\bibitem[{{Tang} {et~al.}(2009{\natexlab{a}}){Tang}, {Ho}, {Girart}, {Rao},
  {Koch}, \& {Lai}}]{tang2009a}
{Tang}, Y.-W., {Ho}, P.~T.~P., {Girart}, J.~M., {et~al.} 2009{\natexlab{a}},
  \apj, 695, 1399

\bibitem[{{Tang} {et~al.}(2009{\natexlab{b}}){Tang}, {Ho}, {Koch}, {Girart},
  {Lai}, \& {Rao}}]{tang2009b}
{Tang}, Y.-W., {Ho}, P.~T.~P., {Koch}, P.~M., {et~al.} 2009{\natexlab{b}},
  \apj, 700, 251

\bibitem[{{Tang} {et~al.}(2013){Tang}, {Ho}, {Koch}, {Guilloteau}, \&
  {Dutrey}}]{tang2013}
{Tang}, Y.-W., {Ho}, P.~T.~P., {Koch}, P.~M., {Guilloteau}, S., \& {Dutrey}, A.
  2013, \apj, 763, 135

\bibitem[{{Tang} {et~al.}(2010){Tang}, {Ho}, {Koch}, \& {Rao}}]{tang2010}
{Tang}, Y.-W., {Ho}, P.~T.~P., {Koch}, P.~M., \& {Rao}, R. 2010, \apj, 717,
  1262

\bibitem[{{Tassis} {et~al.}(2009){Tassis}, {Dowell}, {Hildebrand}, {Kirby}, \&
  {Vaillancourt}}]{tassis2009}
{Tassis}, K., {Dowell}, C.~D., {Hildebrand}, R.~H., {Kirby}, L., \&
  {Vaillancourt}, J.~E. 2009, \mnras, 399, 1681

\bibitem[{{Vall{\'e}e} \& {Fiege}(2006)}]{vallee2006}
{Vall{\'e}e}, J.~P. \& {Fiege}, J.~D. 2006, \apj, 636, 332

\bibitem[{Van~Loo {et~al.}(2014)Van~Loo, Keto, \& Zhang}]{vanloo2014}
Van~Loo, S., Keto, E., \& Zhang, Q. 2014, Submitted to \mnras

\bibitem[{{Wang} {et~al.}(2014)}]{wang2014}
{{Wang}, K., {Zhang}, Q., {Testi}, L., {Tak}, F.~v.~d.,
{Wu}, Y., {Zhang}, H., {Pillai}, T., {Wyrowski}, F.,
{Carey}, S., {Ragan}, S.~E., \& {Henning}, T.}, 2014, \mnras,
439, 3275

\bibitem[{{Wang} {et~al.}(2012){Wang}, {Zhang}, {Wu}, {Li}, \&
  {Zhang}}]{wang2012}
{Wang}, K., {Zhang}, Q., {Wu}, Y., {Li}, H.-b., \& {Zhang}, H. 2012, \apjl,
  745, L30

\bibitem[{{Wang} {et~al.}(2011){Wang}, {Zhang}, {Wu}, \& {Zhang}}]{wang2011}
{Wang}, K., {Zhang}, Q., {Wu}, Y., \& {Zhang}, H. 2011, \apj, 735, 64

\bibitem[{{Wang} {et~al.}(2010){Wang}, {Li}, {Abel}, \& {Nakamura}}]{wang2010}
{Wang}, P., {Li}, Z., {Abel}, T., \& {Nakamura}, F. 2010, \apj, 709, 27

\bibitem[{{Zhang}(2005)}]{zhang2005b}
{Zhang}, Q. 2005, in IAU Symposium, ed. R.~{Cesaroni}, M.~{Felli},
  E.~{Churchwell}, \& M.~{Walmsley}, 135--144

\bibitem[{{Zhang} {et~al.}(2005){Zhang}, {Hunter}, {Brand}, {Sridharan},
  {Cesaroni}, {Molinari}, {Wang}, \& {Kramer}}]{zhang2005a}
{Zhang}, Q., {Hunter}, T.~R., {Brand}, J., {et~al.} 2005, \apj, 625, 864

\bibitem[{{Zhang} {et~al.}(2001){Zhang}, {Hunter}, {Brand}, {Sridharan},
  {Molinari}, {Kramer}, \& {Cesaroni}}]{zhang2001}
---. 2001, \apjl, 552, L167

\bibitem[{{Zhang} {et~al.}(2009){Zhang}, {Wang}, {Pillai}, \&
  {Rathborne}}]{zhang2009}
{Zhang}, Q., {Wang}, Y., {Pillai}, T., \& {Rathborne}, J. 2009, \apj, 696, 268

\end{thebibliography}
\end{document}